\documentclass[11pt]{article}
\usepackage{graphicx} 
\usepackage{amssymb} % For \Box and other math symbols
\usepackage{color}
\usepackage{titlesec}
\usepackage{amsthm}
\usepackage{tabularx}
\usepackage{float}
\usepackage{tikz}
\usepackage{lineno}
\usepackage{hyperref}

\usepackage{setspace}
\usepackage{lipsum} 

\usepackage{amsmath}
\usepackage[
    a4paper,
    left=2.5cm,
    right=2.5cm,
    top=2cm,
    bottom=2cm,
    marginparwidth=0cm,
    marginparsep=0cm,
    footskip=1cm
]{geometry}

\title{From $\mathrm{AdS}_5$ to $\mathrm{AdS}_3$: TsT deformations, Magnetic fields and Holographic RG Flows}
\author{Lucas S. Sousa\textsuperscript{1} \footnote{mail: santos.sousa@unesp.br}}
\begin{document}
\maketitle
\begin{center}
\textsuperscript{1}
Instituto de Fisica Teorica, UNESP-Universidade Estadual Paulista, R. Dr. Bento T. Ferraz 271, Bl. II, Sao Paulo, 01140-070, SP, Brazil
\end{center}
\begin{abstract}
It was previously shown that a D7 brane probe in a D3 brane background with a pure gauge constant magnetic field $\mathrm{B} = \mathrm{H}$ exhibits chiral symmetry breaking and a discrete meson spectrum with Zeeman splitting. In this work, we investigate how these features are modified by a TsT deformation of the background, which renders the Kalb Ramond field physical and radially dependent, thereby obscuring its interpretation as a constant magnetic field.

We show that chiral symmetry breaking persists in the deformed model. The meson spectrum, however, depends on the fluctuation sector. Fluctuations perpendicular to the magnetic field are sensitive to the deformation and, for generic values of the TsT parameter $\mathrm{k}$, do not admit a consistent spectrum due to divergent behavior near the horizon, whereas fluctuations parallel to the magnetic field remain unaffected.

Remarkably, the combined effect of the magnetic field and the TsT deformation singles out the special value $\mathrm{k} = -\frac{1}{\mathrm{H}}$. At this point, the perpendicular modes are restored. Moreover, the Kalb Ramond field becomes constant again, recovering its interpretation as a magnetic field. The resulting effects on the spectrum appear only at order $O(H^2)$, and therefore the Zeeman splitting, if present at all, is shifted to this higher order.

Furthermore, the resulting background from $\mathrm{k} = - \frac{1}{\mathrm{H}}$ is interesting in its own right. The spacetime admits an interpretation in terms of D1 branes and exhibits a degenerate boundary geometry, asymptotically $\mathrm{AdS}_3 \times S^5$, with a degenerate horizon which has nevertheless non-negligeble quantum effects, and then not being completely trustable as a classical solution in that regime. We present a first discussion of the dual field theory interpretation, making connections to D1 and D5 systems, renormalization group flow, defect field theories, and domain wall holography.

\end{abstract}

\newpage

\tableofcontents

\newpage

\section{Introduction}

Since the proposal of the AdS/CFT correspondence by Maldacena \cite{Maldacena1997}, a lot of work has been done in trying to generalize this gauge/gravity duality to a variety of other models, each of them with particular properties. Yet, we hope that, at the end of the day, these models can somehow be useful to approximate results from our Standard Model. In that direction, people have proposed new models and obtained results, generally from the gravity side of the correspondence, to compare with results known to hold, either numerically or analytically, in the Standard Model (SM), or more specifically in Quantum Chromodynamics (QCD). Even if QCD is a theory with fewer symmetries than general SYM models, being neither supersymmetric nor conformal (though QCD is approximately conformal in the UV, where quark masses $m$ are negligible), nor integrable (although in some limits, and by studying amplitudes and form factors, the model shows deep connections with integrability, see \cite{Braun1998}, \cite{Faddeev1995}, \cite{Lipatov1993}), we can still compare certain results with those obtained from gauge/gravity duality. Indeed, while it is an obvious fact that our Standard Model is not $\mathcal{N}=4$ SYM, calculations done in AdS/CFT settings can still produce values close to real results. A famous example is the ratio $\eta/s$ of shear viscosity to entropy density, whose value obtained from holographic calculations in AdS is numerically close to that extracted for the QGP, see \cite{Kovtun2004}.

But a lot of things still need to be implemented, and certainly the supergravity background obtained from $N$ D3 branes alone \cite{Maldacena1997} is not enough to describe all properties we expect, in principle, from a QCD-like gauge theory, namely chiral symmetry breaking, confinement, asymptotic freedom, nontrivial phase transitions at $T \neq 0$ and $\mu \neq 0$, fundamental and antifundamental matter fields, and so on. Each of the models that follow the AdS/CFT correspondence in the direction of reproducing QCD is, in some sense, a deformation of the original model, since the field theory we want to obtain is four dimensional in the end, so we certainly expect a D3 brane configuration, although with more ingredients added to it. Even so, different configurations can provide us with insight into the final gravity background.

For example, \cite{Girardello1999} and \cite{Girardello1998} proposed a way to break conformal symmetry on the gauge side and restore it in the ultraviolet, while at the same time obtaining a confinement-like behavior, \cite{Sonnenschein1999}, \cite{Girardello1999}. It is easy to see the relation between these effects by noting that the Wilson loop is of the area-law type, $W \sim \exp[A]$, which necessarily signals confinement and, consequently, the absence of conformal symmetry. This can be achieved by introducing, from the gravity point of view, a hard wall in the model or by allowing a dilaton-like flow, see \cite{Constable1999}. Regarding mass spectra, several models have made considerable advances in showing how to add flavors, \cite{Karch_2002}, by studying probe branes with $N_c \gg N_f$ in a supergravity background, or how to obtain the glueball mass spectrum, beyond the fact that conformal symmetry is also broken through the introduction of a cutoff, either soft or hard \cite{Karch_2006}, \cite{Csaki1999}. As another example, if we want chiral symmetry breaking, it is necessarily true that supersymmetry must be broken, and the simplest way to do that while generating a quark condensate is to introduce a constant magnetic field into an already established background with probe branes, for example a D7 brane as a probe in the D3 brane geometry. Finally, models simulating chiral symmetry breaking by a condensate have also been proposed in the literature, by introducing black holes and probes \cite{Sakai2004}, \cite{Sakai_20052}, or a constant magnetic field \cite{Filev_2007}, \cite{Alishahiha1999} or even an RG-flow sensitive dilaton that breaks supersymmetry and conformal symmetry, while presenting a confining-like behavior and, by introducing probe branes, chiral symmetry breaking \cite{Constable1999}. We also mention that these problems have been studied from the bottom-up point of view by constructing supergravity or pure gravity models that display the properties we want in order to mimic QCD. This approach goes by the name of AdS/QCD, see \cite{Erlich2005}.

In this paper, we are interested in deforming the model of \cite{Filev_2007}, which is a simple setup consisting of a D3 brane background with D7 brane probes, inspired by \cite{Karch_2002}. The model includes a constant Kalb-Ramond field $\mathrm{B}$, which, as we are going to explain later, can be equivalently interpreted as a constant magnetic field $\mathrm{B}$, and, being in the probe approximation, the model breaks supersymmetry. The magnetic field is responsible for both chiral symmetry breaking and a Zeeman-like effect in the spectrum. We are interested in knowing whether we can weaken the constancy of $\mathrm{B}$, and if so, what the effects of a non constant magnetic field are on the phase structure and on the meson spectrum. To approach this problem, we deform the model by applying TsT transformations in the directions parallel to the constant magnetic field on the supergravity side.

We briefly mention that a magnetic field can be consistently incorporated on the field-theory side without appealing to the probe approximation or interpreting it as a Kalb-Ramond field. This can be achieved by considering a dyonic black hole in the bulk. The AdS$_4$ construction, dual to a $d=3$ condensed-matter system, was presented in \cite{Hartnoll2007}, and this approach was later extended to AdS$_5$ in \cite{DHoker2009}.

\section{Review of constant B background}
\label{review}

This section follows essentially the second section of \cite{Filev_2007}, so familiar reader with the procedure can skip it. 

The near-horizon region of the D3-brane background can be shown \cite{Maldacena1997} to consist of a $AdS_5 \times S^5$ metric, with a self-dual $F_5$ field strength, with a flux over the sphere $S^5$ proportional to $N_c$, and a constant dilaton $\phi$,
{\small \begin{equation}
    \begin{aligned}
        \mathrm{ds}^2 = \frac{u^2}{L^2} ( -dx_0^2 + \sum_{i=1}^3 d x_i^2 ) + \frac{l^2}{u^2} ( du^2 + u^2 d \Omega_5^2 ), \;        F_5 = d C_4 + \dots, \;        \phi = \ln g_s, \;
        L^4 = 4 \pi g_s N \alpha'^2,
    \end{aligned}
\end{equation}}
where $\dots$ are terms imposing the $F_5 = \star F_5$ condition, $L$ is the radius of the space and $\phi$ is the dilaton. Throughout this paper we will adopt $L = 1$. Also, $C_4$ has the form 
\begin{equation}\begin{aligned}
    C_4 = u^4 dx^0 \wedge dx^1 \wedge dx^2 \wedge dx^3.
\end{aligned}\end{equation}
We now embed the D7-brane in the background of the D3-branes. To do that, first we rewrite the radial+sphere part of the metric
\begin{equation}\begin{aligned}
    \mathrm{ds}^2 = \left ( \rho^2 + l^2 \right )  ( -dx_0^2 + \sum_{i=1}^3 d x_i^2 ) + \frac{1}{\rho^2 + l^2} ( d \rho^2 + \rho^2 d \Omega_3^2 + d l^2 + l^2 d \Phi^2 ),
\end{aligned}\end{equation}
assuming $x_0,x_1,x_2,x_3,\rho,d \Omega_3$ to be part of the world-volume of the D7-brane, the embedding is $l = l(\rho)$, $\Phi = \text{ constant}$. Therefore, the pull-back of the metric is given by
\begin{equation}\begin{aligned}
    \mathrm{ds}^2 = \left ( \rho^2 + l^2 \right )  ( -dx_0^2 + \sum_{i=1}^3 d x_i^2 ) + \frac{1}{\rho^2 + l^2} ( \left ( 1 + l'^2 \right ) d \rho^2 + \rho^2 d \Omega_3^2 ).
\end{aligned}\end{equation} 
Moreover, as described in \cite{Filev_2007}, we have to include a constant magnetic field $\mathrm{B}$ with components along $x_2, x_3$,
\begin{equation}\begin{aligned}
    B =\mathrm{H} \; dx_2 \wedge dx_3.
\end{aligned}\end{equation}
We can add such a magnetic field without worrying about whether it is a supergravity solution, because since $\mathrm{H}$ is constant, $\mathrm{B}$ is a pure gauge field and therefore the field strength vanishes, $dB = 0$. Since only the field strength appears in the type IIB supergravity equations of motion, this does not affects the solution. We can now substitute the fields above into the DBI action, also assuming $2\pi = 1$. To understanding why we can purposely confuses the $\mathrm{B}$ field with $\mathrm{F}$, it is clear if we pay attention to the NSNS part of the action, given by
\begin{equation}\begin{aligned}
    S_{NS} = -\mu \int_{M_8} d^8 \xi \sqrt{ -\det (G + B + \alpha' F)},
\end{aligned}\end{equation}
as one can see, $B+F$ appears as an antisymmetric combination, and this shows to be true even in the Wess-Zumino term (more on this later). Consequently, if we only know $A_{23} = B_{23}+F_{23}$, we can say $F_{23}=A_{23}$ is equivalently described by $B_{23}=A_{23}$, with the last being the interpretation of Clifford et. al \cite{Filev_2007}, where if $\mathrm{A}$ is constant, then a constant magnetic field $B_1 = \epsilon_{123} F_{23}$ is equivalent to a constant Kalb-Ramond field $B_{23}$. Then, we can expand to first order in $\alpha'$ by rewriting the above as
\begin{equation}\begin{aligned}
    \sqrt{-\det (G+B + \alpha' F)} = \sqrt{-\det (E + \alpha' F)} = \sqrt{-\det E} \sqrt{\det (1 + \alpha' E^{-1} F)}
\end{aligned}\end{equation}
where $E = G + B$. Making use of the mathematical relation $\det e^{A} = e^{\mathrm{tr}\ln A}$, we approximate $\sqrt{\det(1 + \alpha' \, c_1 )} \approx 1 
+ \frac{1}{2} \alpha' \, \mathrm{tr} c_1  $, 
\begin{equation}\begin{aligned}
    S_{NS} = - \mu \int d^8 \xi \sqrt{-\det E} - \frac{\mu \alpha'}{2} \int \sqrt{-\det E} \mathrm{tr}( E^{-1} F).
\end{aligned}\end{equation}
Moreover, the presence of $F,B$ and $C_4$  requires Wess-Zumino (WZ) terms in the action \cite{Myers1999}.  These terms take the form (ignoring the pullback symbol)
\begin{equation}\begin{aligned}
    S_{WZ} = \sum_i \int C_i \wedge e^{ B + 2 \pi \alpha' F},
\end{aligned}\end{equation}
where $C$ denotes the formal sum of all R–R potentials, and the Wess–Zumino term contains only the combinations that remain linear in $C$ after expanding the exponential. The antisymmetric fields relevant for calculation are
\begin{equation}\begin{aligned}
C_4 ,\quad \tilde C_6 ,\quad B_2 ,\quad F_2.
\end{aligned}\end{equation}
Here, $\tilde C_6$ is an induced charge on D7-brane due to D3, necessary to show consistency (see \cite{Filev_2007} for details). In their embedding one has
\begin{equation}\begin{aligned}
B_2 \wedge B_2 = 0 ,\qquad B_2 \wedge C_4 = 0,
\end{aligned}\end{equation}
so the Wess–Zumino term simplifies. Keeping terms up to first order in $\alpha'$ we obtain
\begin{equation}\begin{aligned}
S_{\mathrm{WZ}} = \mu_7 \int \tilde C_6 \wedge B_2 + \alpha' \mu_7 \int \tilde C_6 \wedge F_2,
\end{aligned}\end{equation}
and as we are going to show later (similar case in \cite{Arean2005}), $\tilde{C}_4$, the magnetic dual of $C_4$, is of order $\alpha'$, and therefore is not present in the action. So, to first order in $\alpha'$ is
\begin{equation}
\begin{aligned}
\label{und}
    S = - \mu \int d^8 \xi \sqrt{-\det E} + \mu \int C_6 \wedge B  - \frac{\mu}{4} \int d^{10} x \sqrt{-G} |d C_6|^2  \\
    - \frac{\mu \alpha'}{2} \int\sqrt{-\det E} \text{ }  \mathrm{tr}( E^{-1} F)
    +  \mu \alpha' \int C_6 \wedge F_2,
\end{aligned}
\end{equation}
with the third term being the kinetic term for $C_6$ in the (super)gravity model. As shown in detail in \cite{Filev_2007}, the equation of motion for the world-volume gauge field $\mathrm{A}$ (with $F = dA$), obtained from the $\alpha'$ terms in the action, induces a constraint on the $C_6$ field. Consequently, the solution of the equation of motion for $C_6$ must satisfy this constraint. The resulting solution is shown to take the form $C_6 = f(\rho, L_0, \psi)$, where $\psi$ is an angle of the $S_5$ sphere.

Once consistency is established for the R–R equation of motion, we can focus on the leading part of the action coming from the NSNS sector, which is, roughly speaking, proportional to $\sqrt{\det E}$. The asymptotic solution for $l(\rho)$, obtained from its own equation of motion, takes the form
\begin{equation}\begin{aligned}
\label{asympl}
    l(\rho) \sim  m + \frac{c}{\rho^2}.
\end{aligned}\end{equation}
As is standard in the holographic literature \cite{Evans_2004}, \cite{Babington_2004}, \cite{Maldacena1997}, the non–normalizable part of a bulk field near the boundary corresponds to the source of the dual operator, while the normalizable part corresponds to its vacuum expectation value. In our case, $m$ plays the role of the quark mass and $c$ is identified with the chiral condensate,
\begin{equation}
\begin{aligned}
    m \sim M, \;   c \sim \langle \bar{\psi} \psi \rangle.
\end{aligned}
\end{equation}
\cite{Filev_2007} showed the relation between $m$ and $c$, and how one can have $m = 0 $ while $c \neq 0$, which represents a chiral symmetry breaking by non-perturbative effects (and not by explicitly breaking it by adding a small mass parameter) \cite{Kruczenski2003},
\begin{equation}
\begin{aligned}
\label{sa1}
   \langle \bar{\psi} \psi \rangle = - \tilde{c} H^2 = \frac{-H^2}{4m }, \; \;   \tilde{c} = \frac{1}{4m }.
\end{aligned}
\end{equation}
It is possible to obtain the meson spectrum of the model by studying the fluctuations of the fields in the supergravity model. These fields are the vector potential $\mathrm{A}$, that represents the gauge field inside the world-volume of the brane, and the scalars $l, \Phi$, representing the perpendicular coordinates to the D7-brane. Moreover, scalar meson spectrum are obtained by fluctuations of the scalar fields and fluctuations of the vector perpendicular to the D3-brane. However, just like was done in \cite{Filev_2007}, we will study the mixing of scalar field fluctuations with the vector part of the vector fluctuation (along the D3-brane).

In \cite{Filev_2007}  the spectrum was obtained for the weak field, $l(p) \sim m + O(H^2)$, from the equation of motion of $A_{0,1,2,3}$ and $\Phi$ ($\Phi = 0 + \Phi (x_0,x_1,\rho) $). They obtained a quantization of the mass to leading order
\begin{equation}\begin{aligned}
    \tilde{M}_0 =2 \sqrt{(n+1)(n+2)},
\end{aligned}\end{equation}
and a splitting by the magnetic field, just like the Zeeman effect,
\begin{equation}\begin{aligned}
    M_{\pm} = M_0 \pm \frac{H}{m},
\end{aligned}\end{equation}
The spectrum was also obtained for strong magnetic field, but as we are going to show later, this is the same for the TsT deformed.

\section{Deformed D3 background}
\label{3}

The TsT transformation was defined in  \cite{Lunin_2005}, \cite{Frolov_2005}, originally in the context of the $\beta$ deformation of $\mathcal{N}=4$ SYM. It has since been applied to a wider class of deformations, including the $\gamma$ deformation, and later proved useful in the study of $T\bar T$ deformations \cite{McGough2016} and their single-trace generalizations \cite{giveon2020tbartlst}. Moreover, the TsT transformation itself has also been extended to a non-abelian deformation, well understood for chiral model, promoting the idea of deforming a torus (to another torus, $\beta$ deformation preserves $U(1)^2$) to a sphere \cite{Lunin2023}, \cite{Lunin2025}. Besides the fact those transformations are are often related to integrable deformations, an important property is that any abelian TsT transformation automatically generates a new supergravity solution, which makes it a particularly powerful tool in holographic constructions.

\subsection*{NSNS fields}
\label{chiral}

We perform a TsT transformation along $x_2$ and $x_3$, the coordinates parallel to the initial $\mathrm{B}$ field. Using the general rules of T duality for bosonic fields \cite{Buscher1987} over $x_2$,
\begin{equation}
    \begin{aligned}
        \tilde g_{yy}=\frac{1}{g_{yy}},\quad
\tilde g_{yi}=\frac{B_{yi}}{g_{yy}},\quad
\tilde g_{ij}=g_{ij}-\frac{g_{yi}g_{yj}-B_{yi}B_{yj}}{g_{yy}},\quad
\tilde B_{yi}=\frac{g_{yi}}{g_{yy}},\\
\quad
\tilde B_{ij}=B_{ij}-\frac{g_{yi}B_{yj}-B_{yi}g_{yj}}{g_{yy}},\quad
\tilde\phi=\phi_0-\tfrac12\ln g_{yy},
    \end{aligned}
\end{equation}
shifting $x_3 \rightarrow x_3 + \lambda x_2$, and then T-``dualizing'' again over $x_2$, we obtain the transformed NSNS fields,
\begin{equation}
\begin{aligned}
\label{deformed}
     G_{22,33} \rightarrow  \frac{l^2 + \rho^2}{ ( 1 + Hk)^2  + k^2 (l^2 + \rho^2)^2} , \;   B_{23} \rightarrow \frac{H + k H^2+  k(l^2 + \rho^2)^2 \, }{( 1 + Hk)^2  + k^2 (l^2 + \rho^2)^2} , \\
\phi \rightarrow \phi_0 + \frac{1}{2} \log \left[ \frac{1}{(1+Hk)^2 + k^2  (l^2 + \rho^2)^2 } \right].
\end{aligned}
\end{equation}
These expressions can be verified, or more efficiently derived, by following the analysis of \cite{Imeroni_2008} or \cite{Catal-Ozer2005}.

\subsection*{RR fields}
\label{secrr}
The previous section focused primarily on evaluating the effects of the TsT deformation on the NSNS sector of the model. However, we still need to consider the RR part of the action. In fact, this sector can introduce constraints, as discussed earlier. Although these terms do not affect chiral symmetry breaking, they play a crucial role in determining the meson spectrum.

As we know, the D3 brane admits a self-dual $F_5$ form \cite{Maldacena1997}, which, together with the $\mathrm{B}$ field, gives rise to a non trivial transformation of the RR fields under TsT and can even generate new fields. Following \cite{Imeroni_2008} and \cite{Meessen1998} for the transformation rules, we obtain
\begin{equation}\begin{aligned}
\label{tst}
   \mathcal{F}_3 = k f^5_{[x_2 x_3]}, \;
    F_5 = f_5 + k [ f_5 \wedge b]_{x_2 x_3} - k [ f_5  ]_{[x_2 x_3]} \wedge B,
\end{aligned}\end{equation}
where $F_5$, $F_3$, and $\mathrm{B}$ are the new, or deformed, fields generated by the transformation, with $\mathrm{B}$ as obtained previously. The notation $[A]_{[x_i x_j]}$ simply indicates a contraction over the coordinates $x_i$ and $x_j$. For a more detailed explanation of \eqref{tst}, see \cite{Imeroni_2008}.

These new fields appear in the Wess Zumino terms of the DBI action, and, as in the undeformed case, we collect the terms that are linear in $C$. Let us emphasize that we are now studying the D7 brane in the deformed supergravity background, where a TsT transformation has been performed along two spatial coordinates. This results in a new $F_3$, suggesting the presence of a new brane, a D1 brane, which couples electrically to $F_3$ and magnetically to $F_7 = -\star F_3$. This configuration is both interesting and highly nontrivial, as it involves multiple coupled RR and NSNS sectors. In the end, we have $F_3$, $F_7$, $F_5$, and also a non zero magnetic flux $H_3 = dB$, with the latter coupled to a brane. 

Before turning to a detailed field interpretation of this configuration, it is useful to present the explicit form of the RR fields generated by the TsT transformation, as these will be required in the fluctuation analysis. We again refer to \cite{Imeroni_2008} for a more detailed discussion. Using the formulas provided there, one finds, since $C_2 = k (c_4)_{2,3}$,
\begin{equation}\label{c4}\begin{aligned}
C_4 = c_4 - k\, (c_4)_{[x_2 x_3]} \wedge B,
\end{aligned}\end{equation}
where $\mathrm{B}$ is the transformed Kalb Ramond field. Thus,
\begin{equation}{
\label{cc4}
    \begin{aligned}
        C_4 = \left ((l^2+\rho^2)^2 - k (l^2+\rho^2)^2 B_{23} \right ) dx_0 \wedge dx_1 \wedge dx_2 \wedge dx_3, \\ 
        \tilde{C}_4 =
\frac{(1+H k)\,l^2\,(l^2 + 2 \rho^2)}{2 (l^2 + \rho^2)^2}\,
\sin(2\psi)  d\psi \wedge d\beta \wedge d\alpha \wedge d\Phi  , \quad
C_2 = k ( l^2 + \rho^2)^2 dx_0 \wedge dx_1 .
    \end{aligned}}
\end{equation}
To obtain $\tilde{C}_4$, with $\star \tilde{f}_5 = d \tilde{C}_4$, one must manipulate the algebra carefully, since the formulas in \cite{Imeroni_2008} do not directly apply to the magnetic duals. The procedure is as follows. Given the $C_2$ form, one has $F_5 = d C_4 +\mathrm{H}\wedge C_2 + \dots$, where the ellipsis indicates the imposition of the self duality condition on $F_5$.

Finally, one can verify from \eqref{cc4} that, in the limit $k \rightarrow 0$, the correct values of $C_4$ and $\tilde{C}_4$ are recovered, see \cite{Filev_2007}. It is worth emphasizing how remarkable this result is in simplifying calculations. Terms involving $C_4$ do not affects the equations of motion, while terms involving $\tilde{C}_4$ do. Even more strikingly, the result for $\tilde{C}_4$ is simply the undeformed value $\tilde{c}_4$ multiplied by a very simple factor $(1 +\mathrm{H}k)$,
\begin{equation}
\label{cc42}
    \begin{aligned}
        \tilde{C}_4 = \tilde{c}_4 (1 +\mathrm{H}k).
    \end{aligned}
\end{equation}
To conclude, we compute $F_5$ from \eqref{cc4}, since if the construction is correct it should at least be self dual. One can verify that this condition is satisfied,
{\small \begin{equation}
    \begin{aligned}
        F_5 =  f_5 + k [ f_5 \wedge b]_{x_2 x_3} - k [ f_5  ]_{[x_2 x_3]} \wedge B = d C_4 +\mathrm{H}\wedge C_2 + d \tilde{C}_4 = \\
        2(1+Hk)\left(\frac{2u^{3}\,du\wedge dx_{0}\wedge dx_{1}\wedge dx_{2}\wedge dx_{3}}{(1+Hk)^{2}+k^{2}u^{4}}-\cos\theta\,\sin^{3}\theta\,\sin(2\psi)\,d\alpha\wedge d\beta\wedge d\theta\wedge d\Phi\wedge d\psi\right),
    \end{aligned}
\end{equation}}
where $u^2 = l^2 + \rho^2$. It is straightforward to check that $\mathcal{F}_5 = \star \mathcal{F}_5$ using the volume form. Moreover, $F_5$ reduces to the D3 brane flux $f_5$ in the limit $k \rightarrow 0$.

\subsection{Chiral Symmetry Breaking}
\label{4}

To study the effects of deforming the model on chiral symmetry breaking, we follow the standard procedure from the gravity side. We solve for the embedding of the probe brane in the background and then examine its asymptotic behavior, as done in \cite{Filev_2007} and reviewed in Section \ref{review}. We work at first order in $\alpha'$, so we only consider the $\sqrt{\det E}$ part of the action, together with the dilaton factor $e^{-\phi}$. As a consequence of the embedding, we assume $l = l(\rho)$ when solving the equations of motion. More explicitly, the action is
\begin{equation}
\begin{aligned}
    S = - \mu \int d^8 \xi \, e^{-\phi(\rho)} \sqrt{-\det E},
\end{aligned}
\end{equation}
where $G$, $B$, and $\phi$ were obtained in \eqref{deformed}. The matrix $E$, apart from the usual terms present in the undeformed case, is given by
\begin{equation}
    \begin{aligned}
        E = \begin{pmatrix}
            \ddots & 0 & 0 & 0 \\
            0 & \frac{ (l^2 + \rho^2)}{ ( 1 + Hk)^2  + k^2 (l^2 + \rho^2)^2} & \frac{\mathrm{H}+ k H^2 + k(l^2 + \rho^2)^2 }{ ( 1 + Hk)^2  + k^2 (l^2 + \rho^2)^2 } & 0 \\
            0 & -\frac{\mathrm{H}+ k H^2 + k(l^2 + \rho^2)^2 }{ ( 1 + Hk)^2  + k^2 (l^2 + \rho^2)^2 } & \frac{ (l^2 + \rho^2)}{ ( 1 + Hk)^2  + k^2 (l^2 + \rho^2)^2} & 0 \\
            0 & 0 & 0 & \ddots
        \end{pmatrix}.
    \end{aligned}
\end{equation}
Rather than writing the equation of motion explicitly, it is more instructive to analyze the quantities $\sqrt{-\det E}$ and $e^{-\phi}$ separately. They are given by
{\small \begin{equation}
    \begin{aligned}
        \sqrt{-\det E} = \frac{ \rho^{3}\sqrt{ H^{2} + (\rho^2 + l(\rho)^2)^2 } \sqrt{ 1 + (l'(\rho))^{2} } }{ (\rho^{2} + l(\rho)^{2}) \sqrt{ (1 +\mathrm{H}k)^{2} + k^2 (\rho^2 + l(\rho)^2)^2 } }, 
        \quad
        e^{-\phi} = \sqrt{ (1 +\mathrm{H}k)^{2} + k^{2} (\rho^{2} + l(\rho)^{2})^{2} },
    \end{aligned}
\end{equation}}
and, as one can explicitly verify, the factor involving $\mathrm{k}$ in the determinant exactly cancels against the factor coming from the dilaton. In other words, the deformed action, or equivalently the deformed Lagrangian $\mathcal{L}_{\text{def}}$, is equal to the undeformed one $\mathcal{L}$ to first order in $\alpha'$. This implies, in particular, that the undeformed and deformed models share the same equation of motion.

This result is remarkable, since the quantities that determine chiral symmetry breaking, namely the coefficients $m$ and $c$, are fixed by the asymptotic behavior of the solution $l(\rho)$ to the equation of motion. Since the equations of motion are identical, the solutions and their asymptotic expansions coincide as well. Therefore, the results are the same as those obtained in \eqref{asympl}, with the relation between $m$ and $c$ given in \eqref{sa1}. Although this outcome may appear surprising at first, it admits a natural interpretation from the field theory point of view, as we are going to show later.

\subsection{Meson spectrum}
\label{5}

To obtain the mesonic spectrum, we need to study the fluctuation solutions of the equations of motion. This requires allowing terms of order $\alpha'^2$ in the action. However, the calculation is rather lengthy, so for the interested reader we refer to Appendix \ref{apa}, while the main algebraic steps are presented in Appendix \ref{apb}.

\subsection*{Weak magnetic field}

We can now study the meson spectrum. In order to analyze the effect of the TsT deformation, we compare our results with those previously obtained for the undeformed case. As in \cite{Filev_2007}, we focus on fluctuations of $\Phi$ to extract the spectrum. As follows from \eqref{phihi} and \eqref{f01}, the corresponding equations must first be decoupled before the spectrum can be analyzed.

A natural strategy would be the following. To study fluctuations perpendicular to the magnetic field plane, $x_{0,1}$, one could first expand the equations to first order in $H$ and $\mathrm{k}$, and then extract the leading correction to the mass spectrum. To access the regime of stronger fields, one would instead need to analyze the spectrum arising from fluctuations along the parallel plane, $x_{2,3}$, since the differential equation governing fluctuations in the $x_{0,1}$ directions becomes highly nontrivial when higher orders are retained.

Before proceeding we emphasize that fluctuations along the $x_{0,1}$ directions are in fact problematic. This is because a term appearing in the corresponding equation of motion diverges in the deformed case as the radial coordinate $u^2 = \rho^2 + L^2 \to 0$, whereas in the undeformed background this divergence is absent. As a consequence, the spectrum associated with these fluctuations is ill-defined and cannot be reliably obtained. This behavior can be understood by noting that deforming the magnetic field through a TsT transformation, for a generic value of the parameter $\mathrm{k}$, effectively introduces a contribution to the background that is independent of the original constant magnetic field. Indeed, from previous results one sees that it is possible to take the limit $H \to 0$ while keeping $B \neq 0$. Intuitively, this can be viewed as adding a background with a non constant $\mathrm{B}$ field on top of the original constant $\mathrm{B}$ field background. Such non constant magnetic backgrounds are, in general, not suitable for defining a well behaved meson spectrum.\footnote{Of course, the dynamics are nonlinear, so this should not be interpreted as a literal superposition, but the is useful.}

To make this issue explicit, we adopt the weak magnetic field approximation, retaining only terms linear in $\mathrm{k}$ and $H$, while neglecting terms of order $H^2$, $k^2$, $kH$, and higher.

\subsubsection*{$ \bullet \quad x_{0,1}$}

From \eqref{f01}, we can manipulate the equation and rewrite it in terms of $F_{01}$, obtaining
{\small \begin{equation}
\begin{aligned}
\label{f01eq}
   \frac{1}{\sqrt{\det E}} e^{\phi_0}\partial_{\rho} \!\left( \frac{\sqrt{\det E}}{G_1 G_3} e^{-\phi_0} \partial_{\rho} F_{01}\right) 
   +\frac{1}{ G_1 G_4} \Delta_{\Omega_3} F_{01} 
   + \frac{1}{G_1^2} \Delta_{0,1} F_{01}  
   + \frac{G_{2}}{G_1 (G_{2}^2 + B^2)}  \Delta_{2,3} F_{01} \\
    -  \frac{e^{\phi_0}}{\sqrt{\det E}} \partial_{\rho} (K(\rho) B_{23})   (\partial_0^2 - \partial_1^2) \Phi  = 0,
\end{aligned}
\end{equation}}
This is only one of the equations we can obtain, since the other equation vanishes as a consequence of the $A_{p,\sigma,\psi,\dots}$ equation \eqref{aperp}. For the scalar $\Phi$, we find
\begin{equation}
\begin{aligned}
\label{phieq}
    \frac{e^{\phi_0} }{\sqrt{\det E} \tilde{C}^{(\partial \phi)^2} } 
    \partial_{\rho} \!\left( e^{-\phi_0} \tilde{C}^{(\partial \phi)^2} \frac{\sqrt{\det E}}{G_3} \partial_{\rho} \Phi \right) 
    +   G_1^{-1}(\rho)\Delta_{0,1} \Phi   
    +  \frac{G_2}{G_2^2 + B^2} \Delta_{2,3} \Phi 
    + G_4^{-1}(\rho) \Delta_{\Omega} \Phi  \\
    - \frac{e^{\phi_0}}{\sqrt{\det E} \tilde{C}^{(\partial \phi)^2}} \partial_{\rho} (K(\rho) B_{23}) F_{01} = 0.
\end{aligned}
\end{equation}
We now have two parameters to work with, $\mathrm{k}$ and $H$, where $\mathrm{k}$ is the deformation parameter and $H$ is the original magnetic field of the undeformed background. Our goal is to understand how the deformation parameter $\mathrm{k}$ modifies the equations. We therefore retain only terms linear in $\mathrm{k}$ and $H$ in the Kalb–Ramond field,
\begin{equation}\begin{aligned}
\label{bex}
    B =\mathrm{H}+ k (\rho^2 + L_0^2)^2.
\end{aligned}\end{equation}
Moreover, to first order we have $K(\rho) = (1+Hk)\tilde{K}(\rho) \approx \tilde{K}(\rho)$, with
\begin{equation}
    \begin{aligned}
        \tilde{K}(\rho) = - R^4 L_0^2  \frac{ 2\rho^2 + L_0^2}{(\rho^2 + L_0^2)^2}.
    \end{aligned}
\end{equation}
The contribution from $e^{\phi_0}$ enters at second order, of order $O(kH)$, and can be neglected at this stage. We also assume $L_0 = m + O(H^2)$, which leads to
{\small \begin{equation}
\begin{aligned}
G_1 = \rho^2 + m^2 , \quad 
G_2 = \rho^2 + m^2 , \quad 
G_3 = \frac{1}{m^2 + \rho^2} , \quad 
G_4 = \frac{\rho^2}{m^2 + \rho^2} , \quad 
C = \frac{m^2}{m^2 + \rho^2} , \quad
\sqrt{\det E} = \rho^3 .
\end{aligned}
\end{equation}}
With these simplifications, the vector equation reduces to
\begin{equation}
\begin{aligned}
    \frac{1}{\rho^3}\partial_{\rho} (\rho^3\partial_{\rho} F_{01})
    +\frac{1}{\rho^2} \Delta_{\Omega_3} F_{01}
    + \frac{1}{(\rho^2 + m^2)^2} \Delta_{0,1} F_{01}
    + \frac{1}{(\rho^2 + m^2)^2}  \Delta_{2,3} F_{01} \\
    -  \frac{1}{\rho^3}\partial_{\rho} (K(\rho) B_{23}) (\partial_0^2 - \partial_1^2) \Phi  = 0,
\end{aligned}
\end{equation}
while the scalar equation becomes
{\small \begin{equation}
\begin{aligned}
    \frac{1}{\rho^3} \partial_{\rho} ( \rho^3  \partial_{\rho} \Phi)
    +  \frac{1}{(\rho^2 + m^2)^2} \Delta_{0,1} \Phi
    +  \frac{1}{(\rho^2 + m^2)^2}\Delta_{2,3} \Phi
    +\frac{1}{\rho^2} \Delta_{\Omega} \Phi
    - \frac{1}{m^2 \rho^3} \partial_{\rho} (K(\rho) B_{23}) F_{01} = 0.
\end{aligned}
\end{equation}}
We can now decouple the system by introducing the combinations
\begin{equation}\begin{aligned}
    \phi_{\pm} = F_{01} \pm m \mathcal{P} \Phi,
\end{aligned}\end{equation}
where $\mathcal{P} = (-\partial_0^2 + \partial_1^2)^{1/2}$. This leads to
{\small \begin{equation}
\begin{aligned}
\label{dif}
   \Bigg(
   \frac{1}{\rho^3} \partial_{\rho} ( \rho^3\partial_{\rho} )
   +  \frac{1}{(\rho^2 + m^2)^2} \Delta_{0,1}
   +  \frac{1}{(\rho^2 + m^2)^2}\Delta_{2,3}
   +\frac{1}{\rho^2} \Delta_{\Omega}
   \mp  \frac{1}{ \rho^3}\partial_{\rho}( K(\rho) B_{23} )\mathcal{P}
   \Bigg) \phi^{\pm} = 0.
\end{aligned}
\end{equation}}
Denoting by $O_1^{k=0}$ the differential operator in \eqref{dif} evaluated at $k=0$, the equation can be written as
\begin{equation}\begin{aligned}
    \left( O_1^{k=0} \pm \frac{4km}{\rho^2} \mathcal{P} \right) \phi^{\pm} = 0.
\end{aligned}\end{equation}
To solve this equation we follow exactly the same ansatz used in \cite{Filev_2007}. Remarkably, the same ansatz remains valid in the present linearized case,
\begin{equation}\begin{aligned}
    \phi_{\pm} = \eta_{\pm}(\rho) e^{-i q_0 x_0 + i q_1 x_1}.
\end{aligned}\end{equation}
The equation then reduces to
\begin{equation}
\begin{aligned}
    \frac{1}{\rho^3} \partial_{\rho} ( \rho^3 \partial_{\rho} \eta_{\pm})
    + \frac{1}{(\rho^2+m^2)^2} M^2_{\pm} \eta_{\pm}
    \mp \frac{4\mathrm{H}m}{(\rho^2 + m^2)^3} M_{\pm} \eta_{\pm}
    \pm \frac{4km}{ \rho^2}M_{\pm}  \eta_{\pm} = 0.
\end{aligned}
\end{equation}
By splitting $\eta_{\pm} = \eta_0 \pm\mathrm{H}\eta_1 \pm k \eta_2$ and $M_{\pm} = M_0 \pm\mathrm{H}M_1 \pm k M_2$, and then adding and subtracting the two differential equations obtained previously, we arrive at three independent differential equations, corresponding to the orders $1$, $H$, and $\mathrm{k}$,
\begin{equation}
\label{eq}
\begin{aligned}
    \frac{1}{\rho^3} \partial_{\rho} ( \rho^3 \partial_{\rho} \eta_0) + \frac{1}{(\rho^2+m^2)^2} M_0^2  \eta_0 = 0, \\
    \frac{H}{\rho^3} \partial_{\rho} ( \rho^3 \partial_{\rho} \eta_1 ) + \frac{1}{(\rho^2+m^2)^2} \left(M_0^2\mathrm{H}\eta_1 + 2\mathrm{H}M_0 M_1 \eta_0 \right)
    - \frac{4\mathrm{H}m}{(\rho^2 + m^2)^3} M_0 \eta_0  = 0, \\
    \frac{k}{\rho^3} \partial_{\rho} ( \rho^3 \partial_{\rho} \eta_2 ) + \frac{1}{(\rho^2+m^2)^2} \left(M_0^2 k \eta_2 + 2 k M_0 M_2 \eta_0 \right)
    + \frac{4km}{ \rho^2}M_0 \eta_0  = 0.
\end{aligned}
\end{equation}
We solve these equations in the order presented. The first equation is standard, and its solution is given in terms of a Gauss hypergeometric function, exactly as expected from \cite{Filev_2007}. Imposing normalizability at infinity leads to a quantization condition on the parameter $M_0$,
\begin{equation}
\begin{aligned}
    \eta_0 = m^{-1 + \frac{\sqrt{m^{2} + M_{0}^{2}}}{m}}
    \left(m^{2} + \rho^{2}\right)^{\frac{m - \sqrt{m^{2} + M_{0}^{2}}}{2m}}
    \times \\
    {}_{2}F_{1}\!\left(
    \frac{m - \sqrt{m^{2} + M_{0}^{2}}}{2m},
    \frac{3}{2} - \frac{\sqrt{m^{2} + M_{0}^{2}}}{2m},
    1 - \frac{\sqrt{m^{2} + M_{0}^{2}}}{m},
    -1 - \frac{\rho^{2}}{m^{2}}
    \right), \\
    \tilde{M}_0 = 2 \sqrt{(n+1)(n+2)} .
\end{aligned}
\end{equation}
To solve the second equation in \eqref{eq}, we substitute the solution for $\eta_0$, specializing to the simplest case $n=0$. Requiring regularity at $\rho = 0$ yields a perturbative solution for $M_1$, which is well behaved,
\begin{equation}
    \begin{aligned}
        M_1 = \frac{1}{m}.
    \end{aligned}
\end{equation}
As a result, the mass spectrum becomes $M_{\pm} = M_0 \pm \frac{H}{m} + \dots$, reproducing the Zeeman splitting discussed in \cite{Filev_2007}.

The difficulty arises in the third equation of \eqref{eq}. Substituting the values of $M_0$ and $M_1$, again for $n=0$, the full solution is highly nontrivial. Instead, we focus directly on the behavior of the equation near $\rho = 0$. In this limit, the asymptotic solution of the differential equation takes the form
\begin{equation}
\label{ilsol}
    \begin{aligned}
        \eta_2 \propto -\frac{2\sqrt{2}\,m^{7/2}\left(m^{3/2}+\sqrt{2m+\sqrt{2}M_{2}}\pi\rho Y_{1}\!\left(\frac{2\sqrt{2m+\sqrt{2}M_{2}}\rho}{m^{3/2}}\right)\right)}{(2m+\sqrt{2}M_{2})\rho^{2}},
    \end{aligned}
\end{equation}
which is ill defined and diverges as $\rho \to 0$. Even if one substitutes $M_2 = - \frac{2m}{\sqrt{2}}$ directly into the differential equation, a divergent contribution remains and cannot be eliminated. This unavoidable divergence originates from the $\sim k/\rho^{3}$ term in the equation of motion, which persists at any order in $H$ and $\mathrm{k}$.

As a consequence, the meson spectrum associated with fluctuations along the $x_{0,1}$ directions is not well defined, since no finite and normalizable solution exists. As anticipated at the beginning of this section, we therefore turn to the analysis of fluctuations along the plane parallel to the magnetic field, where the effects of the TsT deformation turn out to be trivial.

\subsection*{Strong magnetic field}
\label{strong}

To evaluate the effects of the deformation in the strong magnetic field regime, we follow the approach of \cite{Filev_2007} and study fluctuations not in the $x_0,x_1$ plane, but instead along the $x_2,x_3$ directions. This case might be expected to be more interesting, since it is precisely this plane that undergoes the TsT deformation, and one might therefore anticipate nontrivial effects. However, as will become clear from the following analysis, we find a rather surprising result: the TsT deformation has no effect on the meson spectrum associated with fluctuations in the plane parallel to the deformation, even in the strong magnetic field regime. In fact, this statement holds for arbitrary values of $H$ and $\mathrm{k}$, since, as we will show, no expansion is required.

\subsubsection*{$\bullet \quad x_{2,3}$}

Since the TsT deformation along $dx_2^2 + dx_3^2$ preserves the isometries associated with these coordinates, we can Fourier expand the fluctuation $\Phi$ into a radial function, a plane wave, and a spherical harmonic on the $S^3$, following \cite{Filev_2007},
\begin{equation}\begin{aligned}
    \Phi(\rho) = h(\rho) e^{i q x} Y_l(S_3).
\end{aligned}\end{equation}
Substituting this ansatz into \eqref{phieq}, we obtain
{\footnotesize\begin{equation}
\begin{aligned}
    \frac{e^{\phi_0} }{\sqrt{\det E} C^{(\partial \phi)^2} }
    \partial_{\rho} \!\left( \tilde{C}^{(\partial \phi)^2} \frac{\sqrt{\det E}}{G_3} \partial_{\rho} h \right)
    +  \frac{G_2}{G_2^2 + B^2} M^2 h
    + G_4^{-1} l(l+1) h
    - \frac{e^{\phi_0}}{\sqrt{\det E}} \partial_{\rho} (K(\rho) B_{23}) F_{01} = 0,
\end{aligned}
\end{equation}}
while from \eqref{f01eq} we obtain
{\small\begin{equation}
\begin{aligned}
   \frac{1}{\sqrt{\det E}} e^{\phi_0}\partial_{\rho} \!\left( \frac{\sqrt{\det E}}{G_1 G_3} e^{-\phi_0} \partial_{\rho} F_{01}\right)
   +\frac{1}{ G_1 G_4} \Delta_{\Omega_3} F_{01}
   + \frac{1}{G_1^2} \Delta_{0,1} F_{01}
   + \frac{G_{2}}{G_1 (G_{2}^2 + B^2)}  \Delta_{2,3} F_{01} = 0.
\end{aligned}
\end{equation}}
We immediately see that $F_{01} = 0$ is a solution of this sourceless equation. As a result, the remaining equation reduces to a single differential equation for a scalar function depending only on the radial coordinate,
\begin{equation}
\begin{aligned}
\label{lasteq}
    \frac{e^{\phi_0} }{\sqrt{\det E} C^{(\partial \phi)^2} }
    \partial_{\rho} \!\left( e^{-\phi_0} C^{(\partial \phi)^2} \frac{\sqrt{\det E}}{G_3} \partial_{\rho} h \right)
    +  \frac{G_2}{G_2^2 + B^2} M^2 h
    + G_4^{-1} l(l+2) h = 0.
\end{aligned}
\end{equation}
The terms $G_i$ are
\begin{equation}
\begin{aligned}
    G_1 = \rho^2 + l^2, \quad
    G_2 = \frac{\rho^2 + l^2}{(1 +\mathrm{H}k)^2 + k^2 (\rho^2 + l^2)^2}, \quad
    G_3 = \frac{1+l'^2}{\rho^2 + l^2}, \quad
    G_4 = \frac{\rho^2}{\rho^2 + l^2}, \\
        C^{(\partial \phi)^2} = \frac{l^2}{l^2 + \rho^2}.
\end{aligned}
\end{equation}
After substituting these expressions into \eqref{lasteq}, similarly as in the chiral-symmetry case, $e^{-\phi} \sqrt{\det E}$ cancels the $k$ contribution. Moreover, upon algebraic manipulation, each term in the equation becomes independent of the deformation parameter $\mathrm{k}$. Consequently, the final differential equation is exactly identical to the one obtained in \cite{Filev_2007}. This cancellation arises because the TsT deformation enters the metric, the dilaton, and the open string data through the same algebraic combination. As a result, the solution is not affected by the deformation, and the corresponding meson spectrum remains unchanged, and the mass quantization is the one obtained in \cite{Filev_2007},
\begin{equation}
    \begin{aligned}
        M = 2m \sqrt{(n+l+1)(n+l+2)}
    \end{aligned}
\end{equation}

\subsection*{Field theory}

We summarize here the main results of the effect of TsT deforming the model, which has two major implications for the spectrum of the mesonic fluctuations,
\begin{itemize}
    \item The fluctuations over $x_{0,1}$ are sensitive to the deformation, and the TsT completely removes these massive operators.
    \item The fluctuations over $x_{2,3}$ generate the same spectrum at any order, and are insensitive to the TsT deformation.
\end{itemize}

As discussed in the subsection above on chiral symmetry breaking, the TsT effect near the horizon of AdS$_5$ is negligible, in the sense that it reduces to the original model as we take the limit $l^2 + \rho^2 \to 0$, and moreover after a shift $\phi \to \phi +\mathrm{H}k$, which is always allowed. Let's make this more explicitly. In the $\rho^2 + l^2 \to 0 $ limit, the NSNS fields of the background reduces to, after assuming $k$ as a perturbation and $H$ small (which is not necessary, but it is easier to understand the dual field behavior),
\begin{equation}
\begin{aligned}
\label{limit}
      ds &= \frac{ (l^2 + \rho^2)}{( 1 + Hk)^2  + k^2 (l^2 + \rho^2)^2} ( dx_2^2 + dx_3^2)
      \;\longrightarrow\;  (l^2 + \rho^2)( dx_2^2 + dx_3^2) + \dots, \\
    B &= \frac{\big(H + k H^2+  k(l^2 + \rho^2)^2\big) \, dx_2 \wedge dx_3}{( 1 + Hk)^2  + k^2 (l^2 + \rho^2)^2}
    \;\longrightarrow\; (H + k(l^2 + \rho^2)^2) dx_2 \wedge dx_3 + \dots, \\
\phi &=  \phi_0 + \frac{1}{2} \log \left[ \frac{1}{(1+Hk)^2+ k^2  (l^2 + \rho^2)^2 } \right]
\;\longrightarrow\; \phi_0 - \frac{1}{2} k^2 (l^2 + \rho^2)^2 + \dots .
\end{aligned}
\end{equation}

For the UV limit $l^2 + \rho^2 \to \infty$  the situation is less immediate. The NSNS fields behave as
Near the boundary, the same approximation yields
\begin{equation}
\begin{aligned}
\label{limit13}
      ds &\longrightarrow  \frac{1}{k^2(l^2+\rho^2)} ( dx_2^2 + dx_3^2) + \dots, \\
    B &= \frac{\big(H + k H^2+  k(l^2 + \rho^2)^2\big) \, dx_2 \wedge dx_3}{( 1 + Hk)^2  + k^2 (l^2 + \rho^2)^2}
    \;\longrightarrow\; \left (\frac{1}{\mathrm{k}} - \frac{1}{k^3} \frac{1}{u^4} \right ) dx_2 \wedge dx_3 + \dots, \\
\phi &=  \phi_0 + \frac{1}{2} \log \left[ \frac{1}{(1+Hk)^2 + k^2  (l^2 + \rho^2)^2 } \right]
\;\longrightarrow\; \phi_0 + \frac{1}{2} \log \left [ \frac{1}{k^2 ( l^2 + \rho^2)^2} \right] + \dots ,
\end{aligned}
\end{equation}
where $u^2 = l^2 + \rho^2$.

The fact that the near-horizon region of the deformed background coincides with the undeformed background explains why chiral symmetry breaking in the dual field theory is preserved and independent of the deformation parameter $\mathrm{k}$. This is consistent with the expectations from field theory calculations, since chiral symmetry breaking is an infrared phenomenon.

When moving away from the horizon, the small $\mathrm{H}$ and small $\mathrm{k}$ regime shows that the first nontrivial perturbation appears only in the $\mathrm{B}$ field, which ceases to be constant. Consequently, its interpretation as a magnetic field becomes less clear, and one may reconsider the regime in which $\mathrm{B}$ can be consistently interpreted as a constant magnetic field. As shown above, both near the boundary and near the horizon the $\mathrm{B}$ field approaches a constant value. Therefore, in these asymptotic limits the model is expected to describe a constant magnetic field rather than a dynamical Kalb-Ramond field.

\begin{figure}[H]
\centering
\includegraphics[width=1\linewidth]{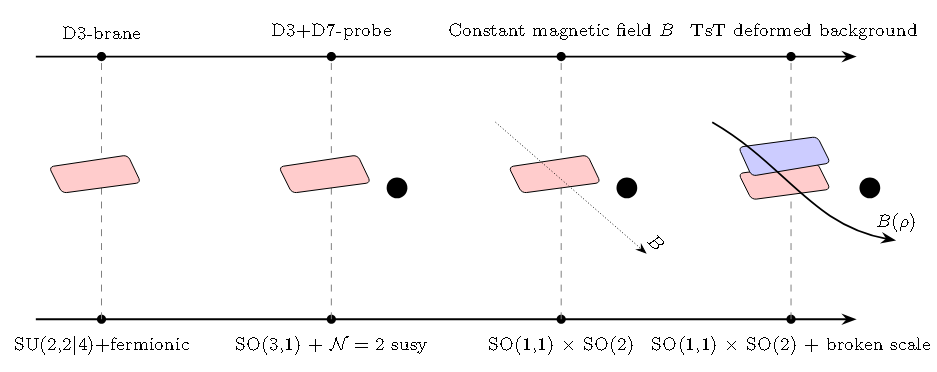}
\caption{Sequence of deformations of the original model and the corresponding loss of symmetries.}
\label{img2}
\end{figure}

Since the metric is already written in Fefferman-Graham coordinates \cite{Fefferman2007}, we can immediately see that the boundary metric is degenerate. Indeed, the metric takes the form (with $z^{-2} = l^2 + \rho^2$)
\begin{equation}
    \begin{aligned}
        \mathrm{ds}^2 = \frac{1}{z^2} \left( dz^2 + g_{ij}(x,z)\, dx^i dx^j \right),
    \end{aligned}
\end{equation}
where, to zeroth order in $z$, we have $g_{22} = g_{33} = 0$. Therefore, the boundary metric $g^0$ is degenerate,
\begin{equation}
    \begin{aligned}
        g^0_{ij}(x) = \mathrm{diag}(1,1,0,0).
    \end{aligned}
\end{equation}
Furthermore, the metric written above resembles a non-isotropic Lifshitz-like geometry \cite{Kachru2008}, though strictly speaking it is not a Lifshitz, since the time $x_0$ scales as the coordinates coordinate $1$, which are the asymptotically the coordinates of the boundary field theory. This can be seen from the invariance of the metric under the anisotropic scaling
\begin{equation}
    \begin{aligned}
    \label{sc}
        v \to \lambda v, \quad x_{0,1} \to \lambda x_{0,1}, \quad x_{2,3} \to \lambda^{-1} x_{2,3}.
    \end{aligned}
\end{equation}
In this case, the dynamical critical exponent is $z=1$\footnote{$z$ is the parameter that determines how the time coordinate scales relative to the spatial coordinates in Lifshitz geometries, where $z \neq 1$.} but the spatial directions $x_{2,3}$ scale differently from $x_{0,1}$. This behavior reveals a hierarchy of symmetry breaking in the model: Lorentz invariance is preserved only in the $(x_{0},x_{1})$ subspace, while scale invariance is broken in a anisotropic way. This chain of broken symmetries is summarized schematically in figure \ref{img2}.

The breaking of the $SO(4)$ symmetry of the D3-brane ($SO(3,1)$ after Wick rotation) down to $SO(2)\times SO(2)$ could, a priori, be attributed to the undeformed model, with $B_{23}$ being responsible for the breaking of the isometry, as illustrated in Fig.\ref{img2} and as explained in \cite{Filev_2007}. Here, however, this isometry breaking acts together with a breaking of scale invariance. As a consequence, while Lorentz symmetry is preserved on the boundary spanned by $x_{0,1}$, with the usual scaling $u,x_{0,1} \to \lambda u, \lambda x_{0,1}$, the fluctuations of the metric along $x_{2,3}$ is invariant under the anisotropic scaling \eqref{sc}. The paper \cite{Cartas2018} has made progress in studying generalizations of this anisotropic behavior as solutions of Einstein-Proca theory, that is, gravity coupled to a massive spin-$1$ vector field, although the present case does not fit directly into the class of solutions studied there.

The $\mathrm{B}$ field breaks the global spacetime symmetry $SO(3,1)$ rather than an internal symmetry. Combined with the anisotropic structure of the background, this makes it difficult to construct an effective UV action that reproduces all expected operators and symmetries. Nevertheless, one may still infer certain general features that such terms must possess from the point of view of the remaining global symmetries. Moreover, the difficulty is not only due to this explicit breaking of spacetime symmetries: the presence of a non-pure gauge and $\rho$-dependent $\mathrm{B}$ field suggests, as conjectured in \cite{Seiberg_1999} and \cite{Lunin_2005}, that the supergravity background may be dual to a non-commutative field theory.

Indeed, \cite{Seiberg_1999} showed that configurations involving a constant Kalb-Ramond field $\mathrm{B}$, and hence open strings ending on D-branes, can be interpreted in terms of a non-commutative field theory, where operators are multiplied using a deformed product. See also \cite{Lunin_2005} and \cite{Frolov_2005} for generalizations to non-constant $\mathrm{B}$ fields and TsT-deformed supergravity backgrounds. In this framework, the usual commutative product of operators is promoted to the non-commutative $\star$-product \cite{Connes_1998},
\begin{equation}
\begin{aligned}
    A(x)\cdot B(x) \;\to\; A(x)\star B(x)
    \;\sim\; A(x)\cdot B(x) + \frac{i}{2}\,\theta^{ij}\,\partial_i A\,\partial_j B ,
\end{aligned}
\end{equation}
and the relevant quantities on the non-commutative gauge theory side are the non-commutativity parameter $\theta$, the open-string metric $G$, and the open-string coupling $g_o$. These are related to the supergravity background through \cite{Seiberg_1999}, \cite{Araujo_2017}, \cite{Araujo_2018}
\begin{equation}
\label{thetatrans}
\begin{aligned}
    G_{ij} = -(2\pi\alpha')^2 (B g^{-1} B)^{ij}, \quad
    \theta^{ij} = (B^{-1})^{ij}, \quad
    g_o = g_s \left(\frac{\det G}{\det g}\right)^{1/4}.
\end{aligned}
\end{equation}
The relation between TsT deformations (or, more generally, deformations of $\mathcal{N}=4$ SYM) and non-commutative field theories has been known for a long time; see, for example, \cite{Bergman2000} and \cite{Dasgupta2000ry}. Moreover, the fact that the boundary metric is degenerate, and therefore exhibits a nontrivial relation to dipole non-commutative field theories, has also been discussed in a similar context in \cite{Bergman2001}. This provides further support for the interpretation presented here, although a more detailed analysis is left for future work.

\begin{figure}[H]
    \centering
    \includegraphics[width=0.8\linewidth]{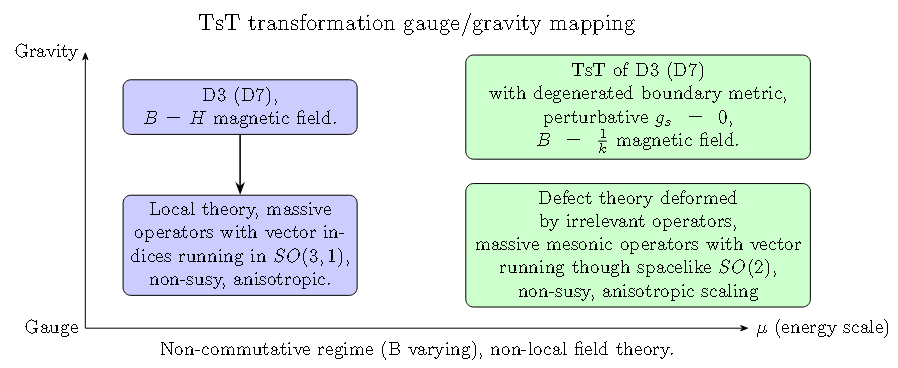}
    \caption{The TsT deforming of the D3-D7 + $\mathrm{H}$ constant \cite{Filev_2007} model in different regime of energies.}
    \label{fig:placeholder}
\end{figure}

We stress one important fact: As discussed, near the horizon the supergravity background reduces to the undeformed case, and this automatically translates to the IR of the field theory being intact by the deformation. Moreover, what is perhaps obvious from the just given information, though is going to be useful later, is that the near-horizon region of this model is not degenerate. Then, the deformed model shares an irrelevant behavior from the RG flow point of view, it is non-local and present defects terms on the Lagrangian since the boundary goes like $\sim \left (\frac{1}{u^2} d \vec{x}_{2,3} + u^2 d \vec{x}_{0,1} \right )$, and therefore the preserved \textit{isometric}\footnote{A symmetry of the metric only, since $\phi(u)$ breaks the scale invariance of $x_{0,1,u}$ at large $u$.} subgroup of the conformal symmetry $\mathrm{SO}(4,2)$ is $\mathrm{SO}(2,2)$. 

Moreover, the coupling is constant asymptotically in both limits, though with different constant values, and $C_{2,4}$ are only known from their constant charge, which translate in the dimension of the gauge group $N_1,N_3$. 

$N_3$ is trivially obtained by integrating $\star F_5$, which in the present background gives
\begin{equation}
    \begin{aligned}
        |N_3| = (1+Hk) \Omega_5,
    \end{aligned}
\end{equation}
the $N_1$, nevertheless, is less trivial. The Page charge can be obtained from the equation of motion,
\begin{equation}
    \begin{aligned}
        d \left (F_7 + B \wedge F_2 \right ) = 0,
    \end{aligned}
\end{equation}
which can be integrate over the specific cycles to obtain $Q_{N_1}$,
\begin{equation}
    \begin{aligned}
       |N_1| = 4 H A_{23} \Omega_5.
    \end{aligned}
\end{equation}
The conserved combination of flux involving $H_3$ is given by $d(e^{-2\phi} H_3 + C_2 \wedge F_5) = 0$. Remarkably, in our background this specific combination of form fields vanishes identically. As a consequence, the associated flux is zero, and the presence of $H = dB$ in the dual field theory manifests itself purely through the non locality of operator interactions.

Therefore, the present supergravity model is dual to a field theory with anisotropic scaling, which is free and coincides with $\mathcal{N}=4$ in the IR. The theory is deformed both by the explicit insertion of defect operators and by deforming existing terms in $\mathcal{L}_{\mathcal{N}=4}$ through their promotion to non local interactions, and it flows asymptotically to a weakly coupled, $g_s \to 0 $, $\mathrm{AdS}_3 \times S^3$ regime. From the energy flow point of view, the gauge group is always $SU(N_1) \times SU(N_3)$, and the internal symmetry $SU(4)$ is preserved in the flow, but supersymmetry is broken and the spacetime symmetry of the model is reduced from $SO(4,2)$ to the isometry group $SO(2,2)$. Even though the $SO(2,2)$ group isometry is suggesting a possible conformal symmetry, the dilaton $\phi(u)$ breaks the scale invariance of the near boundary $\mathrm{AdS}_3$, that would otherwise be present in the IR.

Finally, we emphasize the IR limit of the field theory is precisely that of $d = 4$, $\mathcal{N}=4$ super Yang Mills theory, this flow appears to be driven by defect operators which are irrelevant.

\section{D1 + magnetic field background}

Throughout the whole analysis of the last Section, we have neglected the fact that the theory points to a special value for the parameter $\mathrm{k}$, and therefore we address this issue in the present Section. Indeed, \eqref{cc4}, \eqref{deformed}, and even the explicit form of $K(\rho)$ make it clear that there exists a specific value for the deformation parameter as a function of the magnetic field magnitude,
\begin{equation}
\label{k}
    \begin{aligned}
        \mathrm{k} =  - \frac{1}{\mathrm{H}}.
    \end{aligned}
\end{equation}
Equation \eqref{k} is far from a generic or meaningless result. In fact, one immediately notices that, if \eqref{k} holds, the function $K(\rho)$ vanishes identically. As a consequence, the ill-defined term that appears in the equations of motion for fluctuations perpendicular to the magnetic field disappears. This makes it possible to study these fluctuations consistently again, and potentially to obtain a well-defined meson spectrum in this sector. Moreover, equation \eqref{k} has several additional implications, but for the moment we focus only on the two main physical aspects addressed in this paper, namely chiral symmetry breaking and the hadronic spectrum.

\subsection{Chiral symmetry breaking}

The chiral symmetry breaking calculated in Section \eqref{4} can be straightforwardly repeated here, since we can simply substitute \eqref{k} into the calculations performed there, and the implications are the same as before: the model exhibits chiral symmetry breaking. This time, however, one can directly relate it to the TsT deformation parameter,
\begin{equation}
    \begin{aligned}
        \langle \bar{\psi} \psi \rangle \sim H^2 = \frac{1}{k^2}.
    \end{aligned}
\end{equation}
Then, there is a double interpretation in the present model, since we may interpret the magnetic field as $H = -\frac{1}{\mathrm{k}}$, or instead the TsT parameter as $\mathrm{k} =  -\frac{1}{\mathrm{H}}$. Both points of view are valid, although, following the logic of the calculation, the latter interpretation is more appropriate.

The transformation of the NSNS fields, for the special value \eqref{k}, is
\begin{equation}
\begin{aligned}
     G_{22,33} \rightarrow \frac{H^2}{ (l^2 + \rho^2)} , \quad
     B_{23} \rightarrow -H, \quad
     \phi \rightarrow \phi_0 + \log \left[ \frac{H}{ (l^2 + \rho^2) } \right],
\end{aligned}
\end{equation}
and therefore $\det E$ and $e^{\phi}$ becomes
\begin{equation}
\label{lim1}
    \begin{aligned}
        \sqrt{-\det E} =
       \mathrm{H}\frac{p^{3}\sqrt{\left(H^{2} + (\rho^2 + l(p)^2)^2 \right)}\sqrt{1 + (l'(p))^{2}}}{(p^{2} + l(p)^{2})^2},
        \quad
        e^{-\phi} = \frac{e^{-\phi_0}}{H}(l^2+\rho^2).
    \end{aligned}
\end{equation}

However, one should worry about the divergence of $g_s = e^{\phi}$ in the near horizon scale, as quantum corrections are non-suppressed. Therefore, a classical approach can not be the appropriate regime to be used near the horizon. While this is true, we postpone that discussion for Section~\ref{suap}, and for the moment we just assume this solution is valid even under quantum corrections, in order to study its features.

\subsection{Meson spectrum}

The meson spectrum resulting from fluctuations over $x_{2,3}$ is the same as in the case of Section \ref{strong}, since the overall effects of the TsT deformation cancel among themselves, and therefore we obtain the same result as before.

More interestingly, the fluctuations over the perpendicular plane deserve closer attention. The differential equation for the scalar mesons is the same as in the previous Section, but we rewrite it here for the reader's convenience,
{\small\begin{equation}\begin{aligned}
\label{dif2}
    \frac{e^{\phi_0} }{\sqrt{\det E} \tilde{C}^{(\partial \phi)^2} } 
    \partial_{\rho} \!\left( e^{-\phi_0} \tilde{C}^{(\partial \phi)^2} \frac{\sqrt{\det E}}{G_3} \partial_{\rho} \Phi \right) 
    +   G_1^{-1}(\rho)\Delta_{0,1} \Phi   
    - \frac{e^{\phi_0}}{\sqrt{\det E} \tilde{C}^{(\partial \phi)^2}} \partial_{\rho} (K(\rho) B_{23}) F_{01} = 0.
\end{aligned}\end{equation}}
We recall that
\begin{equation}
    \begin{aligned}
        K(\rho) = (1+Hk)\,\tilde{K}(\rho),
    \end{aligned}
\end{equation}
and therefore \eqref{k} implies $K(\rho) = 0$.

We are going to assume the weak magnetic field limit, which in the present case, one should interpret it as a large $k$ TsT parameter. 

The meson spectrum associated with these fluctuations is straightforward to understand by looking at the differential equation. The linear term in $H$ appearing in \eqref{dif2} vanishes, and therefore the leading correction to the mass spectrum arises only at order $\mathrm{H}^2$. As discussed in the previous Sections, however, this order is technically difficult to analyze, since the embedding function $l(\rho)$ is no longer constant. At this stage, the only robust conclusion we can draw is that, for $\mathrm{k} = -\frac{1}{\mathrm{H}}$, the effects of the deformation on the mass spectrum arising from fluctuations along the $x_{0,1}$ directions are increased to order $\mathrm{H}^2$.

Then, the simply conclusion from the discussion is that, for the mass spectrum over $x_{01}$,
\begin{equation}
    \begin{aligned}
        M_{01} = 2 \sqrt{(n+1)(n+2)}  + O(H^2).
    \end{aligned}
\end{equation}

\subsection{Holographic approach} \label{hol}

For the moment, we will ignore the fact stressed last section regarding the supergravity approximation $1 \gg R^2 \alpha',1 \gg g_s $, and assume the solution is valid in any range of $u \in [0,\infty)$. While we are going to show this is not true, it presents an intuitive interpretation of the scenario. Later, on Subsection \ref{suap} we bring this problem back, and discuss possible ways to overcome this.

To understand the field theory dual to the present supergravity background it is useful to go back to the configuration involving only the D3-brane together with a constant magnetic field. In this case, using the usual coordinates $u^2 = l^2 + \rho^2$, the deformed background takes the form
\begin{equation}
\label{newm}
    \begin{aligned}
        \mathrm{ds}^2 = u^2 ( -dx_0^2 +  dx_1^2 ) + \frac{1}{u^2} \big( \frac{1}{k^2} \left (dx_2^2 + dx_3^2  \right)+  du^2 + u^2 d \Omega_5^2 \big), \quad
        \phi = \phi_0 + \log \frac{H}{u^2}, \\
        B = - H, \quad C_{(2)} = -\frac{u^4}{H},
    \end{aligned}
\end{equation}
where both $C_4$ and its magnetic dual vanish, as can be seen directly from the deformed expressions in \eqref{c4}, and consequently $F_5$ also vanishes. This represents a surprising simplification of the configuration. The vanishing of $C_4$ and $F_5$ means that there is no D3-brane charge present. Therefore, the specific value of the TsT parameter $\mathrm{k}$ in \eqref{k} is special precisely because it removes the original D3-brane content of the background. 

One can understand this perspective by evaluating the Wess-Zumino part of the brane action, as we did in the previous sections for the general $k$ value. Indeed, since \cite{Filev_2007} one can "properly confuses itself" to attribute the value of $\mathcal{F}$ either to $F_{ab}$ or to $B_{ab}$, we can assume $F_{ab} = 0$ (not being too strict about the equation of motion for $F$, as we did before) as $B$ is constant, and then the WZ terms are
\begin{equation}
    \begin{aligned}
        \int_{D_3}  C_4 + C_2 \wedge B + \frac{1}{2} C_0 B \wedge B \subset S_{\mathrm{WZ}}.
    \end{aligned}
\end{equation}
The last term vanishes, since $B^2 = 0$ for the present $B$ field. Then, we see the magnetic field induces a $C_2$ charge in $D_3$, which can be effectively described as a $D_1$ brane when $B = -k^{-1}$.

It is also straightforward to see that the resulting supergravity background does not preserve supersymmetry. Indeed, the dilatino variation reads
\begin{equation}
    \begin{aligned}
        \delta \lambda
        = \frac{1}{2} \left ( \gamma^u \partial_u \phi
        + \frac{e^{\phi}}{2} \, \gamma^{u01} F_{u01} \sigma_1 \right ) \epsilon
        \sim \frac{1}{2} \left ( \gamma^u u^{-1} - \gamma^{u01} u \sigma_1 \right ) \epsilon,
    \end{aligned}
\end{equation}
which does not vanish for any nontrivial Killing spinor $\epsilon$, since $u$ is spacelike and the two term doesn't have the same power of $u$, not allowing for a cancellation of the coefficients. Consequently, this special TsT deformation breaks all the supersymmetry of the original D3-brane background.

The fact that the configuration now contains only a $B_2$ field and a $C_2$ field suggests that the spacetime is sourced by a nontrivial bound state involving D1, F1. To identify the relevant charges more precisely, we can compute the fluxes of the RR fields. The three-form field strength is
\begin{equation}
    \begin{aligned}
        F_3 = -\frac{4 u^3}{H}\;
        dx^0 \wedge dx^1 \wedge du,
    \end{aligned}
\end{equation}
with Hodge dual
\begin{equation}
    \begin{aligned}
       F_7 = - \star F_3
       =
       4 H\,
       \sin^4\psi\,
       \sin^3\beta\,
       \sin\theta\,
       \cos\theta\;
       dx^2 \wedge dx^3
       \wedge d\psi \wedge d\beta \wedge d\alpha
       \wedge d\phi \wedge d\theta.
    \end{aligned}
\end{equation}
The flux of $F_7$, namely $\int \star F_7$, is independent of the radial coordinate and is therefore not useful for charge quantization in this case. On the other hand, $F_3$ leads to a simple, radially independent flux $\int \star F_3$, which can be straightforwardly quantized. We stress, before calculating the flux, that the existence of different definitions for brane charges, some quantized but not gauge-invariant while others the other way around \cite{Marolf2000}, is not something we need to worry in the present background, as $H_3 = 0, F_5 = 0$ and the charges turns out to be the same, $Q_{\text{Max}} = Q_{\text{Page}}$\footnote{We thank prof Carlos Núñez for pointing that.}.

The D1-brane charge is
\begin{equation}
    \begin{aligned}
        Q_{D_1} = - \int \star F_3 = 4 \,\mathrm{H}\, A_{23} \, \Omega_5,
    \end{aligned}
\end{equation}
where $A_{23}$ is the area of $x_{2,3}$ plane with flux through it, and $\Omega_5$ denotes the volume of the five-sphere. We see, moreover, the quantity of $N_1$ is the same as for a generic value of $k$.

In the end, we are left with a D1-brane background together with a constant Kalb-Ramond field, which can again be interpreted as a magnetic field. In this sense, the construction closes consistently, recovering a configuration with a magnetic field but with a completely different origin.

Finally, it is worth emphasizing a subtle but important point about the deformed background. In general, a deformation necessarily preserves some information about the original background, since it is constructed from it. At first sight, it may seem contradictory that the final configuration no longer resembles the original D3-brane geometry. The resolution of this apparent paradox becomes clear once one reuse dimensionful coordinates in the background, where information of the original D3-brane scale reappear implicitly through the deformation parameters.

We recall that we set $L_3=1$ at the beginning of the paper. Restoring dimensions, the background can be written as
{\small
\begin{equation}
    \begin{aligned}
        \mathrm{ds}^2 = \frac{u^2}{L_1^2} ( -dx_0^2 +  dx_1^2 ) + \frac{L_1^2}{k^2 u^2} ( dx_2^2 + dx_3^2 )+ \frac{L_1^2}{u^2} du^2 + L_1^2 d \Omega_5^2 ,\\
        \phi = \phi_0 + \log \frac{H L_1^2}{u^2}, \quad B = \frac{1}{\mathrm{k}}, \quad C_{(2)} = k\frac{u^4}{L_1^4},
    \end{aligned}
\end{equation}}
where $L_1$ is the radius of the deformed background, with dimensions $[L_1]=-1$. We can determine $L_1$ by relating it to the number of D1-branes, obtaining
\begin{equation}
\label{n1}
    \begin{aligned}
         L_1^4 = \Big|\frac{N_1}{4H} \Big| \frac{2 \kappa_{10}^2}{A_{23} \Omega_5} = \Big|\frac{N_1}{4H} \Big| \frac{(2 \pi)^6 g_s^{(0)} \alpha'^3}{A_{23} \Omega_5} ,
    \end{aligned}
\end{equation}
where we are ignoring the volume factor resulting from integrating $S^5$, etc. We also recall that the $S^5$ part of the metric is undeformed, and by consistency we must have
\begin{equation}
\label{n2}
    \begin{aligned}
        L_1 = L_3,
    \end{aligned}
\end{equation}
and we know that,
\begin{equation}
\label{n3}
    \begin{aligned}
        L_3^4 = 4 \pi g_s^{(0)} \alpha'^2 |N_3|^{(\mathrm{k=0})},
    \end{aligned}
\end{equation}
where $\mathrm{k=0}$ is to remind the reader that the flux of $F_5$ was computed before the TsT deformation together with the condition $\mathrm{k=-\frac{1}{\mathrm{H}}}$, and should not be confused with an actual flux in the present background. Combining \eqref{n1}, \eqref{n2}, and \eqref{n3}, we obtain
\begin{equation}
\label{n4}
    \begin{aligned}
        |N_1| = \Big | \frac{16 \pi g_s^{(0)} H A_{23} \Omega_5 \alpha'^2 N_3^{(k=0)}}{(2 \pi)^6 g_s^{(0)} \alpha'^3} \Big | = \frac{16 \pi^4 H  A_{23} N_3}{64 \pi^6 \alpha'} = \frac{1}{4 \pi^2 \alpha'} H A_{23} |N_3^{(k=0)}| \Rightarrow \\
        \frac{N_1}{N_3} =\frac{A_{23}}{4 \pi^2} \frac{H}{ \alpha'}.
    \end{aligned}
\end{equation}
Therefore, the information from the initial background is not completely lost. Equation \eqref{n4} is interesting by itself, since it shows that the number of D1-branes is proportional to the number of D3-branes in the original background. One notice an interesting point, that we are going to discuss more about later: if we compatified $A_{23}$ over a torus with radius $R_1 = R_2 = \sqrt{\alpha'}$, then \eqref{n4} reduces to a simple relation without any numerical factor, $\frac{N_1}{N_3} = H$.

It is interesting to study the different limits of the supergravity background \eqref{newm}, but with caution, since limits in general do not commute. 

\subsubsection*{Near boundary}

Near the boundary, or equivalently in the UV of the field theory, as $u \to \infty$, and writing $g_s = e^{\phi}$, the background behaves as
{\small \begin{equation}
    \begin{aligned}
        \mathrm{ds}^2 \sim u^2 ( -dx_0^2 +  dx_1^2 ) + \frac{du^2}{u^2} + d \Omega_5^2 , \quad
        g_s \to 0 , \quad
        B = \frac{1}{\mathrm{k}}, \quad
        C_2 \to \infty .
    \end{aligned}
\end{equation}}
The metric is effectively eight-dimensional because we have discarded the degenerate directions at the boundary. More precisely, the behavior is similar to what was found in the previous Section, where the boundary metric is degenerate. In the present case, however, since $\mathrm{B}$ is constant, the dual field theory remains local, and there is no need to worry about non-locality associated with non-commutative effects. The degeneracy of the boundary indicates instead that we are deforming the $\mathcal{N}=4$ theory by introducing defect operators. The boundary geometry is $  \mathrm{AdS}_3 \times S^5 .$

Since the $SO(6)$ symmetry is preserved, while the dilaton exhibits a nontrivial flow, one can point out similarities between the present background and Janus-like configurations \cite{Bak2003}, as well as with the Constable–Myers background \cite{Myers1999}.

The fact that the theory is asymptotically $\mathrm{AdS}_3$, together with the vanishing of the coupling, suggests the following UV interpretation:
\begin{equation}
    \boxed{\begin{aligned}
        d = 2 \text{ effective field theory, non-supersymmetric,} \\
        \text{        with } SO(2,2) \text{ global symmetry, }
        \text{gauge group } SU(N_1), \text{ and internal symmetry } SU(4).
    \end{aligned}}
\end{equation}

\subsubsection*{Near horizon}

To study the IR behavior, we move toward the horizon $u \to 0$, where the background becomes
{\small \begin{equation}
    \begin{aligned}
        \mathrm{ds}^2 = \frac{1}{u^2} \big( \frac{1}{k^2} \left (dx_2^2 + dx_3^2  \right) + du^2 + u^2 d \Omega_5^2 \big), \quad
        g_s \to \infty , \quad
        B = \frac{1}{\mathrm{k}}, \quad
        C_2 \to 0 .
    \end{aligned}
\end{equation}}
After rescaling the coordinates $x_{2,3}$, the metric can be brought to a more familiar form. The Kalb-Ramond field is rescaled accordingly but remains constant. In this limit, the coupling diverges, and the field theory becomes strongly coupled.

Near the horizon, the geometry approaches
        $\mathrm{EAdS}_3 \times S^5 $,
where $\mathrm{EAdS}$ denotes Euclidean AdS, or hyperbolic space. Consequently, the IR interpretation of the field theory is
\begin{equation}
    \boxed{\begin{aligned}
        d = 2 \text{ effective field theory, non-supersymmetric,} \\
        \text{with global } SO(3,1), \text{ gauge group } SU(N_1), \text{ and internal symmetry } SU(4).
    \end{aligned}}
\end{equation}
It is interesting how the gauge group and the internal symmetry is preserved in the energy flow. 

What is perhaps more surprising is that, as we explained before, the TsT deformation is in most cases dual to an irrelevant operator. This can be easily seen by taking the limit $k \to 0$ in the deformed supergravity background without imposing $\mathrm{k} = -\frac{1}{\mathrm{H}}$, which results in the undeformed $\mathrm{AdS}_5 \times S^5$, as we have shown. However, if one proceeds in the opposite order and chooses the special value before taking any limit, the deformed contribution to the $x_{2,3}$ part of the metric,
\begin{equation}
    \begin{aligned}
        ds_{\; \;23}^{ \mathrm{TsT}  \;2} = \frac{1}{k^2} \frac{1}{u^2} ( dx_2^2 + dx_3^2),
    \end{aligned}
\end{equation}
is non-perturbative, that is, we can't write $ds_{TsT}^2 \approx \mathrm{ds}^2 + k^2 \delta \mathrm{ds}^2$, which is generally possible by taking the $k \to 0$ or $u \to 0$ limit. Therefore, the field theory is perturbed by operators whose quantum effects shows to be non-perturbative in the IR and in the $k$ parameter. This is why the IR of the field is effectively $x_{0,1}$, since the contributions form the deformation blows up at that energy. At the same time, at UV the theory is effectively $x_{-2,3}$, and not $\mathrm{CFT}_4$. Consequently, it is hard to describe the effective field theory completely in terms of deformed operators in the $\mathcal{N}=4$, but rather it is better to understand it as really a non-isotropic flow.

We emphasize that supersymmetry is not recovered in any asymptotic limit. We recall that the original model was $\mathrm{AdS}_5 \times S^5 + B$, and therefore the deformation acts as a defect in the dual field theory. This happens because the original model lives effectively in $d=10$, while the deformed model becomes degenerate and effectively $d=8$ in certain limits. We stress  that even though the two opposite limits can be described by a $d=8$ model, the region interpolating between those limits is still genuinely $d=10$. Moreover, the effective $d=8$ description involves two distinct sets of non-spherical coordinates, one timelike, $(u,x_0,x_1)$, and the other spacelike, $(u,x_2,x_3)$, but only the first case is truly understandable classically.

\subsection{Analytical-continuation/Wick-Rotation perspective}

Finally, we notice a dual interpretation for the model if we Analytic-Continuate $x_{2,3}$ to complex values, or alternatively Wick-rotate $x_{0,1}$, both limits of the background appear to be $\mathrm{AdS}_3 \times S^5$. Then it is more easy to understand how the field theory behaves as energy flows. We observe that $\mathrm{B}$ can always be gauged away while $C_2$ is known only through its flux. As a result, we can have two possible symmetry interpretations at the asymptotic limits, which do not interpolate as in the previous case,
\begin{equation}
\label{gu}
    \begin{aligned}
        SU(N_1) \text{ gauge group with isometry } \mathrm{AdS}_3 \times S^5 , \\
        SU(i N_1) \text{ gauge group with isometry } \mathrm{EAdS}_3 \times S^5 ,
    \end{aligned}
\end{equation}
since $C_2$ has components over $x_{0,1}$, and the gauge group comes from integrating $\star d C_2$. We can obtain this as a direct consequence of the first case in \eqref{gu}, which is obtained by analytically continuate $x_2 \to i x_2$, while the last is for the continuation $x_0 \to i x_0$. The first case is better to understand the gauge field, while the second is more easier to understand from the supergravity side, as we are going to show.

The appearance of EAdS and continuous groups with ill defined rank, such as a negative or complex value, are concepts present in the domain-wall/cosmology correspondence \cite{McFadden2010}, \cite{McFadden20102}, where amplitudes from a pseudo-QFT (the reason for pseudo is obvious, once you realize the complex rank for the group) in a 3d field theory is obtained by analytically continuation an Euclidean field theory in a Domain-Wal. However, we should mention the analytical continuation in the works cited involves also the radial coordinate, where it is analytically continued to $r \to i t$, such that the RG flow can be understood as a time evolution. Then, to be more precise, at IR we would have after changing $du/u = -dz_1$,
\begin{equation}
    \begin{aligned}
        \mathrm{ds}^2 = L_1^2 \left ( dz_1^2 + e^{2z_1}( dx_2^2 + dx_3^2 ) \right )
    \end{aligned}
\end{equation}
while at UV, changing instead $du/u = dz_2$,
\begin{equation}
    \begin{aligned}
        \mathrm{ds}^2 = L_1^2 \left ( dz_2^2 + e^{2z_2}  ( dx_0^2 + dx_1^2 ) \right ),
    \end{aligned}
\end{equation}
where we absorbed constant conformal factors in $x_{0,1}$ or $x_{2,3}$, since they are not important in the current analysis.

\begin{figure}[H]
    \centering
    \includegraphics[width=0.75\linewidth]{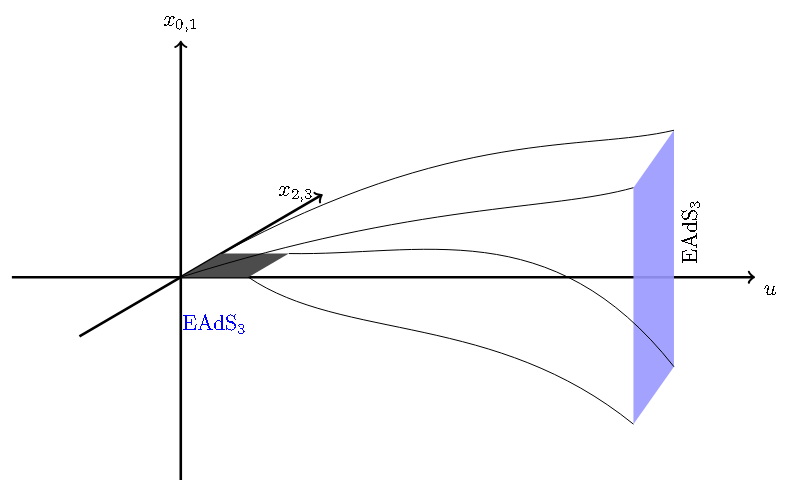}
    \caption{The background as the radius coordinates increases, the planes representing the spacelike domain-walls.}
    \label{fig:i}
\end{figure}

We have a non-isotropic flow, running between two perpendicular domain wall, one over $x_{2,3}$ in the IR, and the other over $x_{0,1}$ in the UV, with the domain-wall being defects on the field theory point of view (which is hard to understand, due to the complexity nature of the gauge group), see Figure \ref{fig:i}.

It is easier to understand the model from the field theory point of view by analyzing the first case of \eqref{gu}, where the QFT is well defined since the group rank is real. In this choice, the asymptotic theories along the RG flow have global symmetry $SO(2,2)$, $d=2$ and $SO(2,2)$, $d=2$, with the dilaton running breaking the conformal symmetry in both limits, and the flow anisotropic. 

\subsection{Supergravity approximation} \label{suap}

Despite the calculations from last sections, one should take the divergence involving $\phi$ more seriously. Indeed, while a divergence of the type $\phi \to -\infty$ is not a problem, as it is equivalent to $g_s \to 0$, a divergence $\phi \to \infty$ is more dangerous, as in that case we are in a strong coupling supergravity, $g_s \to \infty$, regime or even a string theory itself limit.

To verify this, we calculate terms relevant to the deviation of the first order supergravity action. For example, $R$, $R_{\mu  \nu} R^{ \; \mu \nu}$ and $R^{\mu}_{\; \nu \rho \sigma} R_{ \mu}^{\; \; \nu \rho \sigma}$. However, the frame we are working on is the string frame, while those curvature quantities are properly defined in Einstein frame. 

We follow the rules and conventions defined in the nicely written paper \cite{ValeixoBento2025e} to go to the Einstein frame. The transformations are quite simple, since all fields (NS–NS and RR) vanish except for $F_3$ and $\phi$. The Einstein metric can be obtained as
\begin{equation}
    \begin{aligned}
        ds_E^2 = e^{- \frac{\phi-\phi_0}{2}} ds_S^2 = \frac{u}{\sqrt{H} L} \left ( \frac{u^2}{L_1^2} ( -dx_0^2 +  dx_1^2 ) + \frac{H^2 L_1^2}{u^2} ( dx_2^2 + dx_3^2 )+ \frac{L_1^2}{u^2} du^2 + L_1^2 d \Omega_5^2  \right ),
    \end{aligned}
\end{equation}
and since $F_5 = F_1 = 0$, we just need to worry about the transformation of the axion-dilaton $\tau$ and the $G_3$ field, where for $H_3 = d B = 0$ is simply $\tau = i e^{-\phi}, G_3 = F_3$. From \cite{ValeixoBento2025e}, the rules for these fields involves only $\phi_0$\footnote{As explained by the authors of the paper, $\phi_0$ is the vacuum of $\phi$ where the string and Einstein dilaton coincides, then we doesn't need to worry which frame we are for that term.}, which can always be choose to be $0$, and then the mapping is trivial for the present specific case. 

In the Einstein frame the Ricci scalar calculated is $R_E = - \frac{2 \sqrt{H}}{L u} $, but this is expect since it is now a mixing of $R_{S}$ and $\phi_S$. Indeed, when we combine it with $\alpha'_E$ expression, which is \cite{ValeixoBento2025e} $\alpha'_E = e^{-\frac{\phi-\phi_0}{2}} \alpha'_S$, we obtain for the dimensionless scalar involving the curvature the following expressions,
\begin{equation}
    \begin{aligned}
        (R_{\mu \nu} \alpha'_E)^2 = 116 \frac{(\alpha'_S)^2}{L^4}, \quad  (R^{\mu}_{\nu \rho \sigma})^2 (\alpha'_E)^2 = 97\frac{(\alpha'_S)^2}{L^4}, \quad  (R \alpha'_E)^2 = 4\frac{(\alpha'_S)^2}{L^4}.
    \end{aligned}
\end{equation}
Then, if we assume the number of D1-branes $N_1$ (or equivalently, $N_3$) as large enough to suppress the $\alpha'^3$ corrections, the supergravity limit of superstring is still valid, if we properly tune the parameters. 

Nevertheless, the dilaton divergence can't be controlled in that way, and therefore the near-horizon regime of that classical solution is not completely trustable, as it receive quantum corrections in that regime.  Yet, not everything in the near-horizon regime is lost. As is usual in homogeneous dilaton behavior, one can introduce a cutoff in the theory to avoid the quantum regime. Consequently, we would need to introduce a $\Lambda_{\mathrm{IR}}$ cutoff near $u \to 0$, integrating out the horizon and promoting the supergravity solution to be in a box-like spacetime with a mass gap, $r \in [\frac{1}{\Lambda_{IR}}, \infty]$.

Another alternative to introducing an IR cutoff, by taking advantage we are in IIB supergravity, is to perform an S-duality as we flow from large $r$ to the horizon. The S-duality is an isomorphic mapping between two equivalent type IIB supergravity solution, where for the present case we can write the $\mathbb{Z}_2$ transformation as
\begin{equation}
    \begin{aligned}
        \phi \to - \phi, \quad B_2 \to  C_2, \quad C_2 \to B_2.
    \end{aligned}
\end{equation}
Consequently, for $F_3 \neq 0$, we now have a non-zero $H_3$ flux and, as a result, instead of a D1-brane we obtain an F1 string. This new solution admits a classical supergravity regime near the horizon, $u \to 0$, since this limit is now equivalent to $\phi \to -\infty$ and $g_s \to 0$. However, the price we pay is a reversal of the non-classical behavior with respect to the radial coordinate $u$: the problematic regime is now near the boundary, i.e. in the UV. From the field-theory point of view, this way of bypassing quantum corrections is difficult to understand purely in terms of RG flow, as one must interpret it as a flow between two distinct theories. We do not explore this issue further in the present paper, leaving a more detailed discussion for future work.

\subsection{D1/D5 resonances}

Despite these complexities, the model is interesting in its own right. It appears to realize a mixing of different holographic models present in the literature. These include a confining behavior in the IR driven by a dilaton flow \cite{Myers1999}, a defect field theory description \cite{Bak2003}, the presence of a constant magnetic field as in \cite{Filev_2007}, together with group properties and supergravity solutions like those obtained in holographic cosmology \cite{McFadden2010}, \cite{McFadden20102}.

To finish, we discuss one more interesting characteristic of this supergravity solution. As discussed, the metric is
\begin{equation}
    \begin{aligned}
        \mathrm{ds}^2 = u^2 ( -dx_0^2 +  d x_1^2 ) + \frac{1}{u^2} \big( \frac{1}{k^2} \left (dx_2^2 + dx_3^2  \right)+  du^2 + u^2 d \Omega_5^2 \big).
    \end{aligned}
\end{equation}
Since the $S^5$ part of the metric is not affected, we can rewrite the round five sphere metric as a three sphere $S^3$ foliated over $S^1$,
\begin{equation}
\label{d5}
    \begin{aligned}
        d \Omega_5^2 = \cos^2 \theta \, d \Omega_3^2 + \sin^2 \theta \, d \phi^2 + d\theta^2,
    \end{aligned}
\end{equation}
and then, if we approach $\theta \approx 0$ and keep only the leading terms in the metric, together with a change of coordinates in that limit of \eqref{d5} $d\theta^2 + \theta^2 d \phi = dx_4^2 + dx_5^2$, the metric can be written as
\begin{equation}
\label{final}
    \begin{aligned}
        \mathrm{ds}^2 \approx  u^2 ( -dx_0^2 +  d x_1^2 ) + \frac{du^2}{u^2}
        + d \Omega_3^2 + e^{\phi(u)} \widetilde{\mathrm{ds}^2_{M_4}}(u),
    \end{aligned}
\end{equation}
where $\widetilde{\mathrm{ds}^2_{M_4}}$ is a 4-dimensional made of the local flat metric of the compact $S^2$ ($\mathbb{R}^2$) + (rescaled) non-compact ($x_{2,3}$) and dependent on $u$.

This should be compared with the D1/D5 brane background \cite{Brown1986}. The near horizon geometry of that background is $\mathrm{AdS}_3 \times S^3 \times M^4$,
\begin{equation}
\label{or1}
    \begin{aligned}
        \mathrm{ds}^2 = r^2 (dt^2 + dx^2) + \frac{dr^2}{r^2} + d\Omega_3^2
        + \left(\frac{Q_1}{Q_5}\right)^{1/2} \mathrm{ds}^2_{M_4},
    \end{aligned}
\end{equation}
where $M_4$ is a compact internal manifold, and
\begin{equation}
\label{deles1}
    \begin{aligned}
        F_3 = 2 Q_5 \, S^3, \qquad
        e^{2\phi} = \frac{Q_1}{Q_5}, \qquad
        Q_5 = N_5.
    \end{aligned}
\end{equation}
The metric \eqref{or1}, using \eqref{deles1}, can be cast as
\begin{equation}
    \begin{aligned}
        \mathrm{ds}^2 = r^2 (dt^2 + dx^2) + \frac{dr^2}{r^2}
        + d\Omega_3^2 + e^{\phi} \mathrm{ds}^2_{M_4}.
    \end{aligned}
\end{equation}
Despite the similarity between the deformed D1 and D1/D5 background, evidently an exact matching is not possible. The radius $L$, which we set to one in the above calculations, depends on $Q_{1,5}$, which are constant in \eqref{or1} in opposite to our model. Such a dependence would be required by the equations of motion here, and it would be inconsistent to have $L(Q_{1,5})$. Moreover, this limit should be interpreted with some caution, since keeping only the leading terms of $S^5$ near $\theta = 0$ shows to be a inconsistent truncation, since it is not a solution of the type IIB supergravity equations of motion.

Then, perhaps the most appropriate alternative for properly classifying the D1 plus magnetic field background is to compactify the $x_{2,3}$ coordinates by imposing periodicity, so that the metric can be consistently understood as a compact $S^5$ times a warped geometry,
\begin{equation}
    \begin{aligned}
        \mathrm{ds}^2 = \Big (u^2 ( -dx_0^2 +  d x_1^2 ) + \frac{du^2}{u^2} \Big) + e^{\phi(u)} \left ( dx_2^2 + dx_3^2 \right ) + d \Omega_5^2,
    \end{aligned}
\end{equation}
together with the flux
\begin{equation}
    \begin{aligned}
        C_2 = \frac{u^4k }{L_1^4} dx_0 \wedge dx_1.
    \end{aligned}
\end{equation}
It is necessary to impose consistently the $T^2$ periodic identifications on the fields. For NSNS sector it is trivial, since the dilaton $\phi$ is independent of $x_{2,3}$ and the Kalb Ramond field $B$ is constant. Moreover, the modulus of the RR potential $C_3$ is also independent of these coordinates, so periodicity can be imposed in this sector as well. Therefore, the metric becomes
\begin{equation}
\label{inter}
    \begin{aligned}
        \mathrm{ds}^{2 \; \mathrm{T T}}_{\mathrm{k} =  -\frac{1}{\mathrm{H}}} = \underbrace{\mathrm{AdS}_3(u) \times e^{2 \phi(u)} \; T^2}_{\text{Warped Geometry}} \times S^5,
    \end{aligned}
\end{equation}
where $\mathrm{T T}$ in $\mathrm{ds}$ means we compatified $x_{2,3}$ in a 2-torus, $T^2$. \eqref{inter} should be compared with the near-horizon D1/D5 background, $ \mathrm{AdS}_3(u) \times M^4 \times S^3$. 

An appealing feature in favor of the compactification follows from recalling how $N_1$ is obtained, namely from the flux of the dual field $F_7$, which is proportional to the volume of $S^5$ times the area $A_{2,3}$. In the compactified case, $A_{2,3} = \kappa R_2 R_3$, where $R_2$ is the radius of one torus and $R_3$ is the radius of the other, with $\kappa = 4\pi^2$.

\section{Conclusion}
\label{7}

In this paper we generalized the results obtained in \cite{Filev_2007}, where it was shown that a D7-brane probe in a D3-brane background with a constant magnetic field, or pure-gauge Kalb-Ramond field $\mathrm{B}$, not only exhibits chiral symmetry breaking but also a Zeeman splitting in the meson spectrum. Here we relax the assumption of a constant $\mathrm{B}$ field, obscuring the precise equivalence between $F_{12}$ and $B_{12}$, by using the fact that TsT transformations always generate consistent supergravity solutions \cite{Frolov_2005,Lunin_2005}, producing, among other effects, a non-constant Kalb-Ramond field.

More importantly, we showed that chiral symmetry breaking persists in the deformed background. We explained why the condensate $\langle \bar{\psi} \psi \rangle$ depends only on $\mathrm{H}$, the value of $\mathrm{B}$ in the infrared, where it can still be consistently interpreted as a magnetic field. The meson spectrum splits into two distinct sectors again, but only fluctuations parallel to the magnetic field, which are not affected by the TsT deformation, are allowed.

The asymmetric behavior between fluctuations over $x_{0,1}$ and $x_{2,3}$ suggests a non-trivial interpretation from the field theory perspective. The model exhibits not only non-commutative features, but also anisotropic scaling. This interpretation is supported by the presence of a degenerate boundary metric in the gravity solution.

We also studied a special background obtained by the previous one, by fixing the TsT parameter to $\mathrm{k} = -\frac{1}{\mathrm{H}}$, resulting in a pure D1-brane background. In this case, the R-R fields $C_4$ and $F_5$ vanish, which can be interpreted as the disappearance of the original D3-brane. At the same time, the Kalb-Ramond field becomes constant again, $B = -H$, restoring its interpretation as a magnetic field, now with opposite orientation. The first contribution to the meson spectrum of a D7-brane embedded on it is lifted to order $O(H^2)$, generating a non-trivial differential equation which is not evaluate in the present paper, left for future work.

Moreover, the fact that the $\mathcal{O}(\alpha')$ solution for the D7-brane embedding exhibits chiral symmetry breaking is remarkable; however, it is reliable only near the boundary (the UV of the field theory), or at least sufficiently far from the horizon. As discussed in the paper, near the horizon the dilaton dominates and diverges. If one does not introduce an IR cutoff $\Lambda^{-1}_{\mathrm{IR}}$ by hand, higher order terms in the DBI action become non-negligible, overshadowing the physics in that regime.

This D1-brane configuration exhibits both a degenerate boundary and a degenerate horizon of effective dimension $d = 8$, with boundary isometry $\mathrm{AdS}_3 \times S^5$ and the horizon isometry would be $\mathrm{EAdS}_3 \times S^5$ if quantum perturbations were suppressed. One should, therefore, introduce a cutoff before reaching the horizon. The ultraviolet regime of the dual field theory can be interpreted as a non-supersymmetric theory in $d=2$, with gauge group $SU(N_1)$, and internal symmetry $SU(4)$. In the infrared, the gauge field is unclear since the supergravity approximations breaks down, $g_s \gg 1$, but one would naively expect the theory to remains effectively two-dimensional, with spacelike isometry $SO(3,1)$ instead of the $SO(2,2)$ symmetry of the UV. At intermediate energies, however, the field theory is highly non-trivial due to the presence of wrapped directions with a conformal factor $u^{-2}$, which makes a direct analysis difficult.

A more detailed investigation of the dual field theory is left for future work. Since the ultraviolet limit is effectively two-dimensional and $AdS_3$, this background may provide an useful setting to explore irrelevant deformations that are better understood in that dimension, in particular $\mathrm{T}\bar{\mathrm{T}}$-type deformations. Moreover, the effects of introducing an IR cutoff is interesting from the point of view of meson spectrum itself, and one can follow this line to see the implications on the IR-complete D1-brane theory.

The fact that a mixing between the Kalb-Ramond field $B$ and the TsT parameter $k$ generates a single brane background is quite promoting, as it is similar to the Myers effects where a RR-field induces a electrical charge in the brane D$_p$, promoting it to behave effectively like a D$_{p+2}$ \cite{Myers1999}. We doesn't explore it further in the present paper, but is something being done by the author to be published in the future.

\section*{Acknowledgments}

LS would like to thank FAPESP for the support provided under grant 2023/13676-4 , Professor H. Nastase and Professor C Núñez for useful discussions.

\appendix

\renewcommand{\theequation}{\Alph{section}.\arabic{equation}}

\refstepcounter{section}
\setcounter{equation}{0}
\section*{Appendix \Alph{section}}
\label{apa}

\subsection*{Action for fluctuation terms around the classical solution}

We now allow for fluctuations over the $l(p)$ and $\Phi$ (the coordinate, not dilaton) classical solution of the type,
\begin{equation}
\begin{aligned}
\label{fluc}
    l = l_0 + \alpha' \chi, \;    \Phi = \alpha' \Phi.
\end{aligned}
\end{equation}
First we consider the fluctuations \eqref{fluc} on $E$
\begin{equation}\begin{aligned}
    E \rightarrow E_0 + \alpha' \delta_1 + \alpha'^2 \delta_2.
\end{aligned}\end{equation}
Ignoring the explicit form of $\delta_{1,2}$ for while, the NS-NS part of the DBI action is
{\footnotesize\begin{equation}\begin{aligned}
\label{pert}
    \sqrt{\det ( E + \alpha' F)} = \sqrt{\det (E_0+ \alpha' ( F + \delta_1) + \alpha'^2 \delta_2)} = \sqrt{\det E_0} \sqrt{\det (1 + E_0^{-1}(\alpha' ( F + \delta_1) + \alpha'^2 \delta_2))}.
\end{aligned}\end{equation}}
However, we also need to consider fluctuations of the dilaton $\phi$, since it is no longer constant after the TsT deformation. We can write the expansion as
\begin{equation}\begin{aligned}
\label{phi}
    e^{-\phi} = \beta_0 + \beta_1 \alpha'\chi  + \beta_2 \alpha'^2 \chi^2,
\end{aligned}\end{equation}
where $\beta_0 = e^{-\phi} \Big |_{\chi = 0} = e^{-\phi_0}, \beta_1 = \partial_{l_0} e^{-\phi_0}, \beta_2 = \frac{1}{2} \frac{\partial^2 e^{-\phi_0}}{\partial l_0^2}$. To calculate\\
$\sqrt{\det (1 + c_1 \alpha' + c_2\alpha'^2)}$ perturbatively, which is the expression we have in \ref{pert}, we obtain
\begin{equation}
\begin{aligned}
\label{exp}
\sqrt{\det(1 + \alpha'\, c_1 + \alpha'^2\, c_2)} \approx 1 
+ \frac{1}{2} \alpha' \, \mathrm{tr}(c_1) 
+ \alpha'^2 \Bigg( \frac{1}{2} \mathrm{tr}(c_2) - \frac{1}{4} \mathrm{tr}(c_1^2) + \frac{1}{8} (\mathrm{tr}c_1)^2 \Bigg),
\end{aligned}
\end{equation}
and combining with \eqref{phi}, the NSNS part of the DBI action becomes
\begin{equation}
\begin{aligned}
    \int \sqrt{\det E_0} \Big (\beta_0 + \beta_1 \alpha'\chi  + \beta_2 \alpha'^2 \chi^2 \Big )\Big ( 1 + \frac{\alpha'}{2}  \mathrm{tr}( E^{-1}_0 ( F + \delta_1)) +\\
     \frac{\alpha'^2}{8} \left [ 4 \mathrm{tr}( \delta_2) - 2 \mathrm{tr}( E^{-1}_0 ( F + \delta_1) E^{-1}_0 ( F + \delta_1)) + ( \mathrm{tr}( E^{-1}_0 ( F + \delta_1))^2 \right ] \Big ).
\end{aligned}
\end{equation}
The WZ term of the action is, on the other hand,
\begin{equation}
\begin{aligned}
    \int C_6\wedge B + \alpha' \left (C_6 \wedge F_2 + \tilde{C}_4 \wedge B_2 \wedge F_2    \right ) + \frac{\alpha'^2}{2} \left ( C_4 \wedge F_2 \wedge F_2 + C_2 \wedge F_2 \wedge F_2 \wedge B_2 \right ) .
\end{aligned}
\end{equation}
Then, by collecting terms of the full action by $\alpha'$ order,
{\small
\begin{equation}
\begin{aligned}
    \int \\
   \beta_0 \sqrt{\det E_0} + C_6 \wedge B  \\
    + \alpha' \left ( C_6 \wedge F_2  + \sqrt{\det E_0} \beta_1 \chi + \frac{\beta_0}{2}\sqrt{\det E_0} \mathrm{tr}( E^{-1} ( F + \delta_1))  \right) \\
    + \alpha' \tilde{C}_4 \wedge B_2 \wedge F_2 + \alpha'^2 \frac{1}{2}\left ( C_4 \wedge F_2 \wedge F_2 + C_2 \wedge F_2 \wedge F_2 \wedge B_2 \right )\\
    \alpha'^2 \Big[  \beta_0 \left (\sqrt{\det E_0} \frac{1}{8} \left [ 4 \mathrm{tr}( \delta_2) - 2 \mathrm{tr}( E^{-1}_0 ( F + \delta_1) E^{-1}_0 ( F + \delta_1)) + ( \mathrm{tr} E^{-1}_0 ( F + \delta_1))^2 \right ] \right ) \\
    +  \frac{\beta_1 \chi}{2} \sqrt{\det E_0} \mathrm{tr}( E^{-1}_0 ( F + \delta_1)) + \beta_2 \chi^2 \sqrt{\det E_0} \Big ].
\end{aligned}
\end{equation}}
The zeroth order terms were useful to obtain the chiral symmetry breaking and the equation of motion, but we doesn't use them to obtain the fluctuations, and therefore they are not required from now on. Linear terms in the perturbation around the classical solution either vanishes by the stationary action principle or arises as a constraints (first order formalism in $\mathrm{F}$).

We recall that we must check whether this is a consistent background for embedding our D7-brane. This consistency is verified mainly through the equations of motion of the DBI+WZ action, which are essentially the non-abelian gauge field equations of motion for $F_2$, acting as a constraint, together with the $C_6$ equation of motion, as explained in \cite{Filev_2007}. Since we again have a D7 probe, the situation is analogous to the undeformed case, with the only linear term acting on $\mathrm{F}$ being $C_6$. We still have the $\alpha'$ constraint,
\begin{equation}
    \begin{aligned}
    \label{pul}
        \epsilon^{ab   m_1 \dots m_6} \frac{\partial_a C_{m_1 \dots m_6}}{6!} = -\partial_a ( \beta_0 \sqrt{E} E^{[ab]}).
    \end{aligned}
\end{equation}
The right-hand side vanishes for our deformed ansatz because nothing depends on $x_{2,3}$, and therefore we have $d  P[C_6] = 0$. 

At the same time we have the equation of motion resulting from $S_{\mathrm{sugra}} + S_{\mathrm{WZ}}$ with solution near the boundary of the brane
\begin{equation}
    \begin{aligned}
        \partial_L ( \sqrt{-G} d C_6^{L01 \psi \alpha \beta}) = - \frac{\mu_7 \kappa_0^2}{\pi} B(\rho) \delta (L-L_0).
    \end{aligned}
\end{equation}
Since $\partial_x \Theta(x) = \delta (x)$, we can obtain a solution analogous to the one in \cite{Filev_2007}, by calling $C^6_{01 L \psi \alpha \beta} \propto f(\rho)$ such that
\begin{equation}
    \begin{aligned}
    \label{condi}
        f_7^{L01 \rho \psi \gamma \beta} = - \frac{1}{\sqrt{-G}}\frac{\mu_7^2 \kappa_0^2}{\pi} \beta_0 B(p) \Theta(l-l_0),
    \end{aligned}
\end{equation}
where we called $f_7 = d C_6$. We could use the metric to down the indices of $f_7$, and by substituting in the end $k=0$, see the result boils down to the one in \cite{Filev_2007}. 

The rest of the terms at order $\alpha'$ involves only the fields, and its fluctuations, in the zeroth-order equation of motion, and therefore they are the terms that vanishes in the extreme principle of Action. Consequently, we can ignore the $O(\alpha')$ contribution for the action. Therefore, the relevant part of the action for fluctuations are
{\small\begin{equation}
\begin{aligned}
\label{action}
    S = \int \alpha'^2 \sqrt{\det E} (  \beta_2 \chi^2 + \beta_1 \frac{\chi}{2}  \mathrm{tr}( E^{-1} ( F + \delta_1)) + \beta_0  \frac{1}{8} \Big [ 4 \mathrm{tr}( \delta_2) - 2 \mathrm{tr}\Big(  E^{-1} ( F + \delta_1) E^{-1} ( F + \delta_1) \Big) \\
    + ( \mathrm{tr} E^{-1} ( F + \delta_1))^2 \Big ]) + 
     + \alpha' \tilde{C}_4 \wedge B_2 \wedge F_2 + \frac{\alpha'^2}{2} \left ( C_4 \wedge F_2 \wedge F_2 + C_2 \wedge F_2 \wedge F_2 \wedge B_2 \right ).
\end{aligned}
\end{equation}}

\subsubsection*{$\bullet$ NSNS part}

Explicitly, the metric and the $\mathrm{B}$ field we have is
{\scriptsize\begin{equation}
\begin{aligned}
    G_{\text{diagonal}} =  (\rho^2+l^2) \Big( -dt^2  + dx_1^2 + \frac{1}{(1+Hk)^2 + k^2 (\rho^2 + l^2)^2} \left ( dx_2^2 + dx_3^2 \right ) \Big ) 
    + \frac{R^2}{\rho^2 + l^2} ( (1 + l'^2) d\rho^2 + \rho^2 d \Omega_3^2))    \\
    B_{23} = \frac{H + H^2 k + k (\rho^2 + l^2)^2}{((1+Hk)^2 + k^2 (\rho^2 + l^2)^2)} .
\end{aligned}
\end{equation}}
From now on, however, to simplify the presentation of the calculations we adopt the following notation,
\begin{equation} \begin{aligned}
\label{E}
    G = G_1(p) ( -dt^2 + dx_3^2) + G_2(p) (dx_1^2 + dx_2^2) + G_{3}(p) d\rho^2 + G_{4}(p) d \Omega_3^2 ,
\end{aligned} \end{equation}
and therefore, the determinant of $E = G + B$ is
\begin{equation} \begin{aligned}
\label{det}
    \det E = G_1^2 G_2^2 G_4^3 \left (G_3 \det(i) \det(\mu) \det(\Omega) + G_3 B_{23}^2 \det(\mu) \det(\Omega) \right ),
\end{aligned} \end{equation}
while the inverse metric is
\begin{equation} \begin{aligned}
\label{invmet}
    G^{-1} =  G_1(p)^{-1} ( -dt^2 + dx_3^2) + \frac{G_2(p)}{G_2(p)^2 + B_{23}^2} (dx_1^2 + dx_2^2) + G_{3}(p)^{-1} d\rho^2 + G_{4}(p)^{-1} d \Omega_3^2 .
\end{aligned} \end{equation}
Proceeding, since $G$ is diagonal and $\mathrm{B}$ is antisymmetric with only one non-zero component, $B_{23}$, it can be shown \cite{Arean2005} that the inverse of $E$ also satisfies this property and can be decomposed into a diagonal part plus an antisymmetric term with only one non-zero component. That is,
\begin{equation}\begin{aligned}
    E^{-1}_0 = S + J,
\end{aligned}\end{equation}
where $S$ is diagonal and $J$ is antisymmetric, and therefore
\begin{equation}\begin{aligned}
\label{term1}
     ( \mathrm{tr} E^{-1}_0 ( F + \delta_1)) =  \mathrm{tr}S \delta_1 + \mathrm{tr}J F .
\end{aligned}\end{equation}
However, since $\mathrm{B}$ is not constant, the analysis is a bit different form \cite{Filev_2007} because $\delta_1$ is not totally symmetric anymore (because $\partial_l B \neq 0$). In that case, we can split it as
\begin{equation}\begin{aligned}
    \delta_1 = \delta_1^S + \delta_1^A,
\end{aligned}\end{equation}
or explicitly
\begin{equation}\begin{aligned}
\label{delta1}
    \delta^1_{nm} = \underbrace{\left (\frac{\partial G_{nm}}{\partial l}  \Big |_{\chi = 0} \right ) \chi + \frac{1}{\rho^2 + l_0^2} ( \partial_{\rho} l_0 \left( \delta_{mp} \delta_{ni} + \delta_{np} \delta_{mi} \right) \partial_i \chi )}_{\delta_1^S} + \underbrace{\left (\frac{\partial B_{nm}}{\partial l}  \Big |_{\chi = 0} \right ) \chi }_{\delta_1^A}.
\end{aligned}\end{equation}
We can then combine $\delta_1^A + F = \mathcal{F}$ to reduce the term \eqref{term1} to a purely symmetric part plus the product of two antisymmetric matrices
\begin{equation}\begin{aligned}
     \mathrm{tr} E^{-1}_0 ( F + \delta_1) = \mathrm{tr}(S \delta_1^S) + \mathrm{tr}(J \mathcal{F}).
\end{aligned}\end{equation}
Now, for $\mathrm{tr}( E^{-1}_0 ( F + \delta_1) E^{-1}_0 ( F + \delta_1))$, we have
{\footnotesize\begin{equation}
\begin{aligned}
\label{op}
    \mathrm{tr}( E^{-1}_0 ( F + \delta_1) E^{-1}_0 ( F + \delta_1)) =     \mathrm{tr}(S \delta_1^S S \delta_1^S ) + \mathrm{tr}(S \mathcal{F} S \mathcal{F}) + \mathrm{tr}( J \delta_1^S J \delta_1^S ) + 4 \mathrm{tr}( J \delta^1_S S \mathcal{F}) + \mathrm{tr}(J \mathcal{F} J \mathcal{F}),
\end{aligned}
\end{equation}}
and therefore, substituting for $\beta$ in the action,
{\footnotesize\begin{equation}
\begin{aligned}
S =  \int \frac{\sqrt{\det E_0}}{2} \Bigg ( \frac{\partial^2 e^{-\phi_0}}{\partial l_0^2} \chi^2  + \frac{\partial e^{-\phi_0}}{\partial l_0}  (\mathrm{tr}(\delta_1^S S + J \mathcal{F})) \chi+ \, \frac{e^{-\phi_0}}{4} \Bigg[ 
4 \, \mathrm{Tr}(\delta_2) 
- 2 \Big( 
\mathrm{Tr}(S \, \delta_1^S \, S \, \delta_1^S) 
+ \mathrm{Tr}(S \, \mathcal{F} \, S \, \mathcal{F}) 
\\
+ \mathrm{Tr}(J \, \delta_1^S \, J \, \delta_1^S)  + 4 \, \mathrm{Tr}(J \, \delta_1^S \, S \, \mathcal{F}) 
+ \mathrm{Tr}(J \, \mathcal{F} \, J \, \mathcal{F}) 
\Big) + \left( \mathrm{Tr}(S \, \delta_1^S)^2 + 2\mathrm{Tr}(S \, \delta_1^S) \mathrm{Tr}(J \, \mathcal{F}) + \mathrm{Tr}(J \, \mathcal{F})^2 \right)
\Bigg] \Bigg ).
\end{aligned}
\end{equation}}
We can calculate each term explicitly, and this is done in Appendices \ref{termsoftheaction}. There, we show the ``surviving terms'' are the coefficients of the kinetic terms $F^2$, $(\partial \chi)^2$, $(\partial \Phi)^2$, the interaction term $\chi F$, and the ``mass term'' for $\chi$, $\chi^2$. Under these conditions, the NSNS action can therefore be rewritten as
\begin{equation}\begin{aligned}
\label{ac1}
    S = \int \frac{1}{8} ( C_{\chi^2} \chi^2 + C^F_{mnjr} F_{mn} F_{jr} + C^{(\partial x)^2} (\partial \chi)^2 + C^{\chi F}_{nm} \chi F_{nm} + C^{(\partial \phi)^2} (\partial \phi)^2 ),
\end{aligned}\end{equation}
where $\det E$ has been absorbed in the coefficients of the Lagrangian, $C = \sqrt{\det E} \tilde{C}$.

\subsubsection*{$\bullet$ RR part}

The action must now be supplemented by the equations arising from the R-R fields, which, as we have seen, at order $\alpha'^2$ are
\begin{equation}\begin{aligned}
\label{rrr}
    S = \alpha' \tilde{C}_4 \wedge B_2 \wedge F_2 + \frac{\alpha'^2}{2} \left ( C_4 \wedge F_2 \wedge F_2 + C_2 \wedge F_2 \wedge F_2 \wedge B_2 \right ),
\end{aligned}\end{equation}
since the new fields $C_2$ is only present at $\alpha'^2$, and $D_6, C_6$ is only present at $\alpha'$, the results of the calculations are the same as in \cite{Filev_2007} but with a term $(1+Hk)$ multiplying it, as explained in Section \ref{secrr}, 
\begin{equation}
\tilde{C}^4  - \frac{\alpha'}{g_s} \, \frac{\sin 2\psi}{2} \, K(\rho) \, \partial_a \Phi \, d\psi \wedge d\gamma \wedge d\beta \wedge dx^a
\end{equation}
where,
\begin{equation}
K(\rho) = - R^4 L_0^2\frac{2 \rho^2 + L_0^2}{(\rho^2 + L_0^2)^2} ( 1 + Hk).
\end{equation}
In \cite{Filev_2007} they used the Bianchi identity $dF = 0$ together with the fact that $\partial B = 0$. The last is no longer true, however. But this is not a problem with we keep $K(\rho) B$ together, as we are going to show. The first term of \eqref{rrr}, for $F = F_{mn}\, dx^m \wedge dx^n$, is
{\footnotesize\begin{equation}
\begin{aligned}
    - \int \frac{\alpha'^2}{g_s} \, \frac{\sin 2\psi}{2} \, K(\rho) \, \partial_a \Phi B_{23} F_{mn} \, d\psi \wedge d\alpha \wedge d\beta \wedge dx^2 \wedge dx^3  \wedge dx^a \wedge dx^m \wedge dx^n = \\
    -\int  \frac{\alpha'^2}{g_s} \, \frac{\sin 2\psi}{2} \, \, \Phi \Big (\partial_a( K(\rho) B_{23})F_{mn}  + K(\rho) B_{23} \partial_a   F_{mn} \Big) \, d\psi \wedge d\alpha \wedge d\beta \wedge dx^2 \wedge dx^3  \wedge dx^a \wedge dx^m \wedge dx^n,
\end{aligned}
\end{equation}
}
and then, using the Bianchi identity for $\mathrm{F}$ in the last term, $\partial_{[a} F_{mn]}  = 0$, and that $\mathrm{k}$ and $\mathrm{B}$ only depends on $\rho$, we obtain
\begin{equation}
\begin{aligned}
         -        \frac{\alpha'^2}{g_s} \, \frac{\sin 2\psi}{2} \, \, \Phi \partial_{\rho} ( K(\rho) B_{23})F_{01}  \, d\psi \wedge d\alpha \wedge d\beta \wedge dx^2 \wedge dx^3  \wedge d \rho \wedge dx^0 \wedge dx^1 = \\
         \int d^8 \xi \frac{\alpha'^2}{g_s} \, \frac{\sin 2\psi}{2} \,  \Phi \partial_{\rho} ( K(\rho) B_{23})F_{01} .
\end{aligned}
\end{equation}
The second and third term from \eqref{rrr} can be treated together since they are of order $F^2$. Moreover, because $C_4$ and $C_2 \wedge B_2$ have components along the D3-brane world-volume, the structures involving these terms are identical, and we can write it as
\begin{equation}\begin{aligned}
\label{abcd}
    \frac{\alpha'^2}{8 g_s} \int d^8 \Xi \chi(p) F_{ab} F_{cd} \epsilon^{abcd},
\end{aligned}\end{equation}
where we defined $\Xi(p) = C^4_{0123}(p) + C_{01}(p) B_{23}(p)$. Also, the indices in \eqref{abcd} are over transverse coordinates to D3-brane, $\gamma, \psi, \gamma, \phi$. The action is
\begin{equation}\begin{aligned}
    S_{RR} = \frac{\alpha'^2}{8 g_s} \int d^8 \xi \left [\chi(p) F_{ab} F_{cd} \epsilon^{abcd} -2\sin 2 \psi \Phi \partial_{\rho} (K(p) B_{23}) \delta_{0a} \delta_{1b} F_{ab}  \right ],
\end{aligned}\end{equation}
which contributes for the $\Phi$ equation and for $dA$. Notice that the first term above only contributes if index $a=\rho,\psi,\beta,\gamma$ and the second only if $a = 0,1$.

\subsection*{Equation of motion of the fluctuations}

We now calculate the equation of motion for the fluctuations. The interested reader can look at the appendices \ref{apb}, where all calculations are open and made explicitly.

\subsubsection*{$\bullet \quad \chi$}

The $\chi$ equation of motion from  \eqref{ac1} is 
\begin{equation}\begin{aligned}
& 2 C_{\chi^2} \chi +   C^{\chi F}_{nm} F_{nm} 
- \partial_i \Big( 2 C^{(\partial \chi)^2} \, \partial_i \chi  \Big) = 0 
\end{aligned}\end{equation}
with
\begin{equation}\begin{aligned}
    C = \sqrt{(\det E)} \tilde{C},
\end{aligned}\end{equation}
and $\tilde{C}$ was obtained in Appendix \ref{termsoftheaction}.

\subsubsection*{$\bullet \quad \Phi$ }

The equation of motion is derived in detail in the appendices \ref{secphi}. There, we start with
\begin{equation}\begin{aligned}
\frac{1}{4} \partial_i \Big( C^{(\partial \Phi)^2} \, \partial_i \phi \Big) -\frac{\sin 2 \psi}{4}  \partial_{\rho} (K(p) B_{23}) \delta_{0a} \delta_{1b} F_{ab}= 0 ,
\end{aligned}\end{equation}
and, after some manipulation, we can write is more appropriately,
\begin{equation}
\begin{aligned}
    \partial_{\rho} ( \sqrt{\det E} \tilde{C}^{(\partial \phi)^2} G_3^{-1} \partial_{\rho} \Phi)       \\
    +\sqrt{\det E} \tilde{C}^{(\partial \phi)^2} \left ( G_1^{-1}(p)\Delta_{\eta} \Phi   +  \frac{G_2(p)}{G_2(p)^2 + B_{23}^2} \Delta_{\delta} \Phi + G_4(p)^{-1} \Delta_{\Omega} \Phi \right ) -\\
     \sin 2 \psi \partial_{\rho} (K(p) B_{23}) F_{01} = 0.
\end{aligned}
\end{equation}
Notice that the equations are almost the same as in the original paper from \cite{Filev_2007}, but there $\sqrt{\det E} $ is called $g$, and $\mathrm{B}$ appears always outside $\partial (K B_{23})$. Also, the $\Delta_{M_j}$ is the d'Alambertian/Laplacian in the $M_j$ space, so
\begin{equation}
    \begin{aligned}
        \Delta_{\eta} = - \partial_0^2 + \partial_1^2, \; \;
        \Delta_{\delta} = \partial_2^2 + \partial_3^2, \\
        \Delta_{\Omega_3} = \; \; \; \text{Laplacian in spherical harmonic}.
    \end{aligned}
\end{equation}
Finally, we rewrite it as,
{\small\begin{equation}
\begin{aligned}
\label{phihi}
    \frac{1}{\sqrt{\det E}}\partial_{\rho} ( \sqrt{\det E} \tilde{C}^{(\partial \phi)^2} G_3^{-1} \partial_{\rho} \Phi) +    \tilde{C}^{(\partial \phi)^2} \left ( G_1^{-1}(p)\Delta_{\eta} \Phi   +  \frac{G_2(p)}{G_2(p)^2 + B_{23}^2} \Delta_{\delta} \Phi + G_4(p)^{-1} \Delta_{\Omega} \Phi \right ) \\
    - \frac{\sin 2 \psi}{\sqrt{\det E}} \partial_{\rho} (K(p) B_{23}) F_{01} = 0.
\end{aligned}
\end{equation}}

\subsection*{$\bullet \quad A_s$}

First, we want to impose the constraints $A_{\mu} = 0$, where $\mu$ is for coordinates not into the D3-brane world-volume (so $\mu = \psi, \beta, \gamma, \rho$ ) and $i$ is an index for the whole spacetime. Moreover, we gauge fix $\partial \cdot A = 0$.

\subsubsection*{$\bullet \quad s \in \psi, \beta, \gamma, \rho$}

The details of the derivation are in \ref{secaper}. The equation of motion for the perpendicular coordinates of the vector to the brane are
\begin{equation}\begin{aligned}
\label{aperp}
    S^{rr} \partial_s \partial_r A_r = 0,
\end{aligned}\end{equation}
which can be  seen as imposing the following constraints in the vector components along the world-volume of the D3-brane,
\begin{equation}\begin{aligned}
    \partial_1 A_1 - \partial_0 A_0 = \partial_2 A_2 + \partial_3 A_3 = 0.
\end{aligned}\end{equation}
These are exactly the same conditions as in \cite{Filev_2007}.

\subsubsection*{$\bullet \quad s \in 2,3$}

Again, the details are in \ref{seca23}. We obtain
{\small\begin{equation}
\begin{aligned}
    \partial_{\rho} ( G^{pp}  \sqrt{\det E}e^{-\phi_0}    S_{pp} S_{ss}\partial_{\rho} A_s) +   \sqrt{\det E} e^{-\phi_0}  \frac{G_{2} G_4^{-1}}{G_{2}^2+B^2} \Delta_{\Omega_3} A_s  +  \sqrt{\det E}e^{-\phi_0}\frac{G_{2} G_1^{-1}}{G_{2}^2+B^2} \Delta_{0,1} A_s  \\
    +  \sqrt{\det E} e^{-\phi_0}\frac{G_{2}}{G_{2}^2+B^2} \Big (\frac{G_2}{G_2 + B^2}  \Delta_{1,2} A_s \Big) + \partial_r ( C^{\chi F}_{rs} \chi) = 0,
\end{aligned}
\end{equation}}
but since $G_{pp} = S^{-1}_{pp}$,
\begin{equation}
\begin{aligned}
    \frac{1}{\sqrt{\det E}}\partial_{\rho} (   e^{-\phi_0} \frac{G_2}{G_2^2 + B^2} \partial_{\rho} A_s) +   e^{-\phi_0}  \frac{G_{2} G_4^{-1}}{G_{2}^2+B^2} \Delta_{\Omega_3} A_s  + e^{-\phi_0}\frac{G_{2} G_1^{-1}}{G_{2}^2+B^2} \Delta_{0,1} A_s  \\
    +  e^{-\phi_0}\frac{G_{2}}{G_{2}^2+B^2} \Big (\frac{G_2}{G_2 + B^2}  \Delta_{1,2} A_s \Big) + \frac{1}{\sqrt{\det E}}\partial_r ( C^{\chi F}_{rs} \chi) = 0.
\end{aligned}
\end{equation}

\subsubsection*{$\bullet \quad s \in 0,1$}

The details are in \ref{sec01},
\begin{equation}
\begin{aligned}
    \partial_{\rho} ( G^{pp}  \sqrt{\det E}e^{-\phi_0}    S_{pp} S_{ss}\partial_{\rho} A_s) + \sqrt{\det E} e^{-\phi_0}  \frac{1}{G_1 G_4} \Delta_{\Omega_3} A_s  + \sqrt{\det E} e^{-\phi_0}  \frac{1}{G_1^2} \Delta_{0,1} A_s \\
+    \sqrt{\det E}  e^{-\phi_0} \frac{1}{G_1} \Big (\frac{G_{2}^2}{G_{2}^2 + B^2}  \Delta_{1,2} A_s \Big) + \partial_r \Bigg ( \Bigg (\partial_{\rho} (K(p) B_{23}) (\delta_{0r} \delta_{1s} - \delta_{0s} \delta_{1r}) \sin 2 \psi \Bigg ) \Phi \Bigg  ) = 0.
\end{aligned}
\end{equation}
So we have the same case as before,
\begin{equation}
\begin{aligned}
\label{f01}
    \frac{1}{\sqrt{\det E}} \partial_{\rho} (   e^{-\phi_0}\frac{1}{G_{1}} \partial_{\rho} A_s) + e^{-\phi_0}  \frac{1}{G_1 G_4} \Delta_{\Omega_3} A_s  +  e^{-\phi_0}  \frac{1}{G_1^2} \Delta_{0,1} A_s  \\ + e^{-\phi_0} \frac{1}{G_1} \Big (\frac{G_{2}^2}{G_{2}^2 + B^2}  \Delta_{1,2} A_s \Big) + \frac{1}{\sqrt{\det E}} \partial_r \Bigg (\partial_{\rho} (K(p) B_{23}) (\delta_{0r} \delta_{1s} - \delta_{0s} \delta_{1r}) \sin 2 \psi  \Phi \Bigg  ) = 0
\end{aligned}
\end{equation}

\refstepcounter{section}
\setcounter{equation}{0}
\section*{Appendix \Alph{section}}
\label{apb}

\subsection*{Terms of the action}
\label{termsoftheaction}
We give the entries value of each matrix here ( notice that $\mathrm{tr}(E) = \mathrm{tr}(G+B) = \mathrm{tr}(G)$). Explicitly, $\delta_2$ is
{\footnotesize\begin{equation}\begin{aligned}
    \delta^2_{mn} = \frac{1}{2} \frac{\partial^2 E_{mn}}{\partial l^2} \; \; \bigg |_{\chi = 0} \; \;   \chi^2 + \frac{1}{\rho^2 + l_0^2}( (\partial_m \chi )(\partial_n \chi)  + l^2\partial_m \Phi \partial_n \Phi)    -  \frac{2 l_0 l_0'}{(\rho^2 + l_0^2)^2}\left( \delta_{mp} \delta_{ni} + \delta_{np} \delta_{mi} \right)  \chi \partial_i \chi .
\end{aligned}\end{equation}}
Then,
\begin{equation}
\begin{aligned}
    \mathrm{tr}( \delta_2 ) = \left(\frac{1}{2} \frac{\partial^2 \mathrm{tr}(G)}{\partial l_0^2}  \right ) \chi^2 + \frac{1}{\rho^2 + l_0^2} \Big ( (\partial_i \chi)^2 + l_0^2(\partial_i\Phi)^2 \Big )   -  \frac{4 l_0 l_0'}{(\rho^2 + l_0^2)^2} \chi \partial_{\rho} \chi,   \\
    S = S_{mn},    J = J_{mn},    \mathcal{F} = F_{mn} + \delta^1_A = F_{mn} + \frac{\partial B_{mn}}{\partial l_0}   \chi, \\
    \delta^1_S = \left (\frac{\partial G_{nm}}{\partial l_0}  \right ) \chi + \frac{1}{\rho^2 + l_0^2} \Big(  l_0'\left( \delta_{mp} \delta_{ni} + \delta_{np} \delta_{mi} \right) \partial_i \chi  \Big),
\end{aligned}
\end{equation}
with $l_0' = \partial_{\rho} l_0$. Let's take each term of \eqref{action} separately, ignoring the $\sqrt{\det E}$ factor,
\begin{equation} \begin{aligned}
    \frac{\partial^2 e^{-\phi_0}}{\partial l_0^2} \chi^2
\end{aligned} \end{equation}
and
{\small\begin{equation} \begin{aligned}
 \frac{\partial e^{-\phi_0}}{\partial l_0}  (\mathrm{tr}(\delta_1^S S + J \mathcal{F})) \chi =  \frac{\partial e^{-\phi_0}}{\partial l_0}   \Bigg (\Big (\left (\frac{\partial G_{mn}}{\partial l_0}  \right ) \chi + \frac{1}{\rho^2 + l_0^2} \Big(  l_0'\left( \delta_{mp} \delta_{ni} + \delta_{np} \delta_{mi} \right) \partial_i \chi  \Big) \Big ) S_{nm}\\
 + J_{nm} ( F_{mn} + \frac{\partial B_{mn}}{\partial l_0} \chi) \Bigg ) \chi,
\end{aligned}\end{equation}}
\begin{equation} \begin{aligned}
     \mathrm{Tr}(J \, \mathcal{F})^2  = J_{nm} \Big ( F_{mn} + \frac{\partial B_{mn}}{\partial l_0}   \chi  \Big) J_{jr} \Big (  F_{rj} + \frac{\partial B_{rj}}{\partial l_0}   \chi \Big),
\end{aligned} \end{equation}
{\footnotesize\begin{equation} \begin{aligned}
   2 \mathrm{tr}(S \delta_1^S) \mathrm{tr}( J F) = 2 S_{mn} \Big (  \left (\frac{\partial G_{nm}}{\partial l_0}  \right ) \chi + \frac{1}{\rho^2 + l_0^2} \Big(  l_0'\left( \delta_{mp} \delta_{ni} + \delta_{np} \delta_{mi} \right) \partial_i \chi  \Big) \Big ) J_{jr} \Big (  F_{rj} + \frac{\partial B_{rj}}{\partial l_0}   \chi \Big ),
\end{aligned} \end{equation}}
\begin{equation} \begin{aligned}
    \mathrm{tr}( S \delta_1^S)^2 = S_{mn} \Big (  \left (\frac{\partial G_{nm}}{\partial l_0}  \right ) \chi + \frac{1}{\rho^2 + l_0^2} \Big(  l_0'\left( \delta_{mp} \delta_{ni} + \delta_{np} \delta_{mi} \right) \partial_i \chi  \Big) \Big )  \\
    \times S_{jr} \Big (  \left (\frac{\partial G_{rj}}{\partial l_0}  \right ) \chi + \frac{1}{\rho^2 + l_0^2} \Big(  l_0'\left( \delta_{rp} \delta_{jk} + \delta_{jp} \delta_{rk} \right) \partial_k \chi  \Big) \Big),
\end{aligned} \end{equation}
\begin{equation} \begin{aligned}
    \mathrm{tr}( J \mathcal{F} J \mathcal{F} ) = J_{rm} \Big ( F_{mn} + \frac{\partial B_{mn}}{\partial l_0}   \chi \Big ) J_{nj} \Big ( F_{jr} + \frac{\partial B_{jr}}{\partial l_0}   \chi \Big ),
\end{aligned} \end{equation}
{\small\begin{equation} \begin{aligned}
    4 \mathrm{tr}( J \delta_1^S S \mathcal{F} ) = 4 J_{rn} \Big (\left (\frac{\partial G_{nm}}{\partial l_0}  \right ) \chi + \frac{1}{\rho^2 + l_0^2} \Big(  l_0'\left( \delta_{mp} \delta_{ni} + \delta_{np} \delta_{mi} \right) \partial_i \chi  \Big) \Big) S_{mj} \Big ( F_{jr} + \frac{\partial B_{jr}}{\partial l_0}   \chi \Big ),
\end{aligned} \end{equation}}
\begin{equation} 
\begin{aligned}
    \mathrm{tr}(J \delta^S_1 J \delta^S_1) &= J_{rn} \Big ( \left (\frac{\partial G_{nm}}{\partial l_0}  \right ) \chi + \frac{1}{\rho^2 + l_0^2} \Big(  l_0'\left( \delta_{mp} \delta_{ni} + \delta_{np} \delta_{mi} \right) \partial_i \chi  \Big) \Big ) \\
    &\quad \times J_{mj} \Big ( \left (\frac{\partial G_{jr}}{\partial l_0}  \right ) \chi + \frac{1}{\rho^2 + l_0^2} \Big(  l_0'\left( \delta_{jp} \delta_{rk} + \delta_{rp} \delta_{jk} \right) \partial_k \chi  \Big) \Big ),
\end{aligned} 
\end{equation}
\begin{equation}
\begin{aligned}
    \mathrm{tr}(S \delta_1^S S \delta_1^S) &= S_{rn} \Big (\left (\frac{\partial G_{nm}}{\partial l_0}  \right ) \chi + \frac{1}{\rho^2 + l_0^2} \Big(  l_0'\left( \delta_{mp} \delta_{ni} + \delta_{np} \delta_{mi} \right) \partial_i \chi  \Big) \Big) \\
    &\quad \times S_{mj} \Big (\left (\frac{\partial G_{jr}}{\partial l_0}  \right ) \chi + \frac{1}{\rho^2 + l_0^2} \Big(  l_0'\left( \delta_{jp} \delta_{rk} + \delta_{rp} \delta_{jk} \right) \partial_k \chi  \Big) \Big) ,
\end{aligned} 
\end{equation}
\begin{equation} \begin{aligned}
    \mathrm{tr}(S \mathcal{F} S \mathcal{F}) = S_{rm} \Big(  F_{mn} + \frac{\partial B_{mn}}{\partial l_0}   \chi  \Big ) S_{nj}\Big(  F_{jr} + \frac{\partial B_{jr}}{\partial l_0}   \chi \Big),
\end{aligned} \end{equation}
\begin{equation}
\begin{aligned}
   4    \mathrm{tr}( \delta_2 )  =  2\left(\frac{\partial^2 \mathrm{tr}(G)}{\partial l_0^2}  \right ) \chi^2 + \frac{4}{\rho^2 + l_0^2} \Big ( (\partial_i \chi)^2 + l_0^2(\partial_i\Phi)^2 \Big )   -  \frac{l_0 l_0'}{(\rho^2 + l_0^2)^2} \chi \partial_{\rho} \chi .
\end{aligned}
\end{equation}
Finally, collecting term by term, we obtain

\subsubsection*{$ \bullet \; \chi \partial \chi$}

\begin{equation}
\begin{aligned}
\Bigg (\frac{\partial e^{-\phi_0}}{\partial l_0} \frac{1}{\rho^2 + l_0^2} l_0' 2 S_{pi}
 + \frac{e^{-\phi_0}}{4} \Bigg ( -\frac{8 l_0 l_0'}{(\rho^2+l_0^2)^2}\delta_{ip} + 2 \frac{l_0'}{\rho^2 + l_0^2} \,2 S_{pi}\, J_{jr} \, \frac{\partial B_{rj}}{\partial l_0} \, \\
+\frac{l_0'}{\rho^2 + l_0^2} \, 2 S_{pi} \, S_{jr} \, \frac{\partial G_{rj}}{\partial l_0} \,  + \frac{l_0'}{\rho^2 + l_0^2} \, S_{mn} \, \frac{\partial G_{nm}}{\partial l_0} 2 S_{pi} \,  \\
 -8 \frac{l_0'}{\rho^2 + l_0^2} \, (S_{pj} J_{ri} + S_{ij}J_{rp})  \, \frac{\partial B_{jr}}{\partial l_0} \, -2 \Big ( 
\frac{l_0'}{\rho^2 + l_0^2} \,  (J_{jp} J_{ri} + J_{pi} J_{ij})  \, \frac{\partial G_{jr}}{\partial l_0} \,\\
 + \frac{l_0'}{\rho^2 + l_0^2} \,  \, \frac{\partial G_{nm}}{\partial l_0} \, (J_{mp} J_{in} + J_{pn} J_{mi}) \,  \,  \Big ) \\
-2 \Big (
\frac{l_0'}{\rho^2 + l_0^2} \,  (S_{ri} S_{pj} + S_{ij} S_{rp})  \, \frac{\partial G_{jr}}{\partial l_0} \,+ \frac{l_0'}{\rho^2 + l_0^2} \, \frac{\partial G_{nm}}{\partial l_0} \, (S_{in} S_{mp} + S_{pn} S_{mi}) \, \Big ) \Bigg ) \Bigg) \chi \partial_i \chi. 
\end{aligned}
\end{equation}
Or, since $S$ and $G$ is diagonal and $\partial_{\rho} \chi = 0$, and $\mathrm{B}$ and $J$ has only $2,3$ non-zero components,
\begin{equation}
\begin{aligned}
\Bigg (\frac{\partial e^{-\phi_0}}{\partial l_0} \frac{1}{\rho^2 + l_0^2} l_0' 2 S_{pp}
 + \frac{e^{-\phi_0}}{4} \Bigg (-\frac{8 l_0 l_0'}{(\rho^2+l_0^2)^2} + 4 \frac{l_0'}{\rho^2 + l_0^2} \, S_{pp}\, J_{jr} \, \frac{\partial B_{rj}}{\partial l_0} \,  \\
+\frac{l_0'}{\rho^2 + l_0^2} \,4 S_{pp} \, S_{jr} \, \frac{\partial G_{rj}}{\partial l_0} \,    \\
-8 \Big (
\frac{l_0'}{\rho^2 + l_0^2} \,  S_{pp} S_{pp}   \, \frac{\partial G_{pp}}{\partial l_0} \, \Big ) \Bigg ) \Bigg) \chi \partial_{\rho} \chi. 
\end{aligned}
\end{equation}
Therefore, only $\chi \partial_{\rho} \chi$ is non-zero and present. We can, then call $\chi \partial_{\rho} \chi = 2^{-1} \partial_{\rho} \chi^2$, integrate by parts and then add the term to the $\chi^2$ coefficient term, reducing  to
\begin{equation}
\begin{aligned}
\chi \partial_i \chi = 0.
\end{aligned}
\end{equation}
We collect equally each term,

\subsubsection*{$ \bullet \; \chi^2$}

{\small\begin{equation}
\begin{aligned}
    \Big ( \frac{\partial^2 e^{-\phi_0}}{\partial l_0^2}  + \frac{\partial e^{-\phi_0}}{\partial l_0} \Big (\frac{\partial G_{mn}}{\partial l_0}S_{nm}  + J_{nm} \frac{\partial B_{mn}}{\partial l_0} \Big) \\
    +\frac{e^{-\phi_0}}{4} \Big [ J_{nm} \frac{\partial B_{mn}}{\partial l_0} \, J_{jr} \frac{\partial B_{rj}}{\partial l_0} \, + 2 \, S_{mn} \frac{\partial G_{nm}}{\partial l_0} \, J_{jr} \frac{\partial B_{rj}}{\partial l_0} \,  + S_{mn} \frac{\partial G_{nm}}{\partial l_0} \, S_{jr} \frac{\partial G_{rj}}{\partial l_0} \,  + 2 \, \frac{\partial^2 G_{mm}}{\partial l_0^2}   \\
    -2 \Big (  
    J_{rm} \frac{\partial B_{mn}}{\partial l_0} \, J_{nj} \frac{\partial B_{jr}}{\partial l_0} + 4 \, J_{rn} \frac{\partial G_{nm}}{\partial l_0} \, S_{mj} \frac{\partial B_{jr}}{\partial l_0} \, + J_{rn} \frac{\partial G_{nm}}{\partial l_0} \, J_{mj} \frac{\partial G_{jr}}{\partial l_0} \,  + S_{rn} \frac{\partial G_{nm}}{\partial l_0} \, S_{mj} \frac{\partial G_{jr}}{\partial l_0} \, \\
    + S_{rm} \frac{\partial B_{mn}}{\partial l_0} \, S_{nj} \frac{\partial B_{jr}}{\partial l_0} \, 
 \Big ) \Big] \Big) \chi^2,
 \end{aligned}
\end{equation}}

\subsubsection*{$ \bullet \; F^2$}

\begin{equation}\begin{aligned}
\frac{e^{-\phi_0}}{4} \Big (J_{nm}  \, J_{rj}  - \Big ( J_{rm} J_{nj} - J_{jm} J_{nr} + S_{rm} S_{nj} - S_{jm} S_{nr} \Big) \Big ) F_{mn} F_{jr},
\end{aligned}\end{equation}

\subsubsection*{$ \bullet \; (\partial \chi)^2$}

\begin{equation}
\begin{aligned}
e^{-\phi_0} \Big (\frac{ 1}{\rho^2+l_0^2}\,\delta_{ik}  + \frac{ l_0'^2}{(\rho^2 + l_0^2)^2}  S_{pp} S_{ki} \Big)(\partial_i \chi)(\partial_k \chi) ,
\end{aligned}
\end{equation}

\subsubsection*{$ \bullet \; F \partial \chi$}

{\footnotesize\begin{equation}
\begin{aligned}
\frac{e^{-\phi_0}}{4} \Big (2 \frac{l_0'}{\rho^2 + l_0^2} \, S_{mn} (\delta_{mp}\delta_{ni} + \delta_{np}\delta_{mi}) \, J_{jr} \, -4 \frac{l_0'}{\rho^2 + l_0^2} \, (\delta_{mp}\delta_{ni} + \delta_{np}\delta_{mi}) \, (S_{mr}J_{jn} - S_{mj} J_{rn} )\,\Big )\, F_{rj} \, \partial_i \chi
\\
=\frac{e^{-\phi_0}}{4} \frac{2 l_0'}{\rho^2 + l_0^2} \Big ( S_{pi} J_{jr} + S_{ip} J_{jr} - 2 (S_{pr} J_{ji} - S_{pj} J_{ri} + S_{ir} J_{jp} - S_{ij} J_{rp}  \Big ) \Big )  F_{rj} \, \partial_i \chi \\
=\frac{e^{-\phi_0}}{4} \frac{4 l_0'}{\rho^2 + l_0^2} \Big ( S_{pi} J_{jr}  - (S_{pr} J_{ji} - S_{pj} J_{ri} + S_{ir} J_{jp} - S_{ij} J_{rp}  \Big ) \Big )  F_{rj} \, \partial_i \chi \\
=\frac{e^{-\phi_0}}{4} \frac{4 l_0'}{\rho^2 + l_0^2} \Big ( S_{pi} J_{jr}  - S_{pr} J_{ji} +S_{pj} J_{ri}   \Big )  F_{rj} \, \partial_i \chi \\
 =\frac{e^{-\phi_0}}{4} \frac{4 l_0'}{\rho^2 + l_0^2} \Big ( S_{pi} J_{jr}  + S_{pr} J_{ij} +S_{pj} J_{ri}   \Big ) F_{rj} \, \partial_i \chi,
\end{aligned}
\end{equation}}
where we used that $J_{p \dots} = 0$. The term in parenthesis is cyclic in $i,j,r$, so we can partial integrate and obtain
\begin{equation}\begin{aligned}
\label{pfdx}
     -\frac{e^{-\phi_0} l_0'}{\rho^2 + l_0^2} \partial_i \Big ( S_{pi} J_{jr}  + S_{pr} J_{ij} +S_{pj} J_{ri}   \Big ) F_{rj} \,\chi.
\end{aligned}\end{equation}
Therefore, this is really a $F \chi$ term. So we can also say $C^{F \partial \chi} = 0$
\subsubsection*{$ \bullet \; \chi F$}

{\tiny
\begin{equation}
\begin{aligned}
\Big (  \frac{\partial e^{-\phi_0}}{\partial l_0}  J_{jr} + \frac{e^{-\phi_0}}{2} \Big ( \,J_{nm}\left(\frac{\partial B_{mn}}{\partial l_0}\right)\,J_{rj} + \,S_{mn}\left(\frac{\partial G_{nm}}{\partial l_0}\right)\,J_{rj} \\
-  \Big ( \,\left(\frac{\partial B_{mn}}{\partial l_0}\right)\,(J_{rm} J_{nj} - J_{jm} J_{nr})+ 2\,\left(\frac{\partial G_{nm}}{\partial l_0}\right)\, (J_{rn} S_{mj} -J_{jn} S_{mr})+ \left(\frac{\partial B_{mn}}{\partial l_0}\right)\,(S_{nj}S_{rm} - S_{nr}S_{jm}) \Big ) \Big) \Big )\chi F_{jr}.
\end{aligned}
\end{equation}
}
Then, we can add \eqref{pfdx}
\begin{equation}
\begin{aligned}
\Big (  \frac{\partial e^{-\phi_0}}{\partial l_0}  J_{jr} + \frac{e^{-\phi_0}}{2} \Big ( \,J_{nm}\left(\frac{\partial B_{mn}}{\partial l_0}\right)\,J_{rj} + \,S_{mn}\left(\frac{\partial G_{nm}}{\partial l_0}\right)\,J_{rj} \\
-  \Big ( \,\left(\frac{\partial B_{mn}}{\partial l_0}\right)\,(J_{rm} J_{nj} - J_{jm} J_{nr})+ 2\,\left(\frac{\partial G_{nm}}{\partial l_0}\right)\, (J_{rn} S_{mj} -J_{jn} S_{mr}) \\
+ \left(\frac{\partial B_{mn}}{\partial l_0}\right)\,(S_{nj}S_{rm} - S_{nr}S_{jm}) \Big )  
+\frac{2 l_0'}{\rho^2 + l_0^2} \partial_i \Big ( S_{pi} J_{jr}  + S_{pr} J_{ij} +S_{pj} J_{ri} \Big)  \Big ) \Big) F_{jr} \,\chi,
\end{aligned}
\end{equation}
simplifying further,
\begin{equation}
\begin{aligned}
\Big (  \frac{\partial e^{-\phi_0}}{\partial l_0}  J_{jr} + \frac{e^{-\phi_0}}{2} \Big ( \,J_{nm}\left(\frac{\partial B_{mn}}{\partial l_0}\right)\,J_{rj} + \,S_{mn}\left(\frac{\partial G_{nm}}{\partial l_0}\right)\,J_{rj} \\
-  \Big ( \,2\left(\frac{\partial B_{mn}}{\partial l_0}\right)\,J_{rm} J_{nj} + 2\,\left(\frac{\partial G_{nm}}{\partial l_0}\right)\, (J_{rn} S_{mj} -J_{jn} S_{mr})+ 2 \left(\frac{\partial B_{mn}}{\partial l_0}\right)\,S_{nj}S_{rm} \Big )  \\
+\frac{2 l_0'}{\rho^2 + l_0^2} \partial_i \Big ( S_{pi} J_{jr}  + S_{pr} J_{ij} +S_{pj} J_{ri} \Big)  \Big ) \Big) F_{jr} \,\chi.
\end{aligned}
\end{equation}

\subsubsection*{$ \bullet \; (\partial \Phi)^2$}

\begin{equation}\begin{aligned}
    \frac{e^{-\phi_0}}{4} \frac{4}{\rho^2 + l_0^2} l_0^2 (\partial_i \Phi)^2.
\end{aligned}\end{equation}

\subsection*{Equation of motion}

\subsubsection*{$\bullet \; \phi$}
\label{secphi}

\begin{equation} \begin{aligned}
\frac{1}{4} \partial_i \Big( C^{(\partial \phi)^2} \, \partial_i \phi \Big) -\frac{\sin 2 \psi}{4}  \partial_{\rho} (K(p) B_{23}) \delta_{0a} \delta_{1b} F_{ab}= 0 
\end{aligned} \end{equation}
or, rewritten more appropriately,
\begin{equation} \begin{aligned}
    \partial_i ( \sqrt{\det E} \tilde{C}^{(\partial \phi)^2} G^{ij} \partial_j \Phi) - \sin 2 \psi\partial_{\rho} (K(p) B_{23}) F_{01} = 0.
\end{aligned} \end{equation}
Now, $C^{(\partial \phi)^2}$ only depends on $\rho$. We can split the first term as following,
\begin{equation} \begin{aligned}
     \partial_i ( \sqrt{\det E} \tilde{C}^{(\partial \phi)^2} G^{ij} \partial_j \Phi) =  \partial_{\rho} ( \sqrt{\det E} \tilde{C}^{(\partial \phi)^2} G^{pp} \partial_{\rho} \Phi) +  \partial_{x_i} ( \sqrt{\det E} \tilde{C}^{(\partial \phi)^2} G^{x_ix_j} \partial_{x_j} \Phi) \\
     +  \partial_{\mu} ( \sqrt{\det E} \tilde{C}^{(\partial \phi)^2} G^{\mu \nu} \partial_{\nu} \Phi) +  \partial_{\alpha_i} ( \sqrt{\det E} \tilde{C}^{(\partial \phi)^2} G^{\alpha_i \alpha_j} \partial_{\alpha_j} \Phi).
\end{aligned} \end{equation}
Using \eqref{E}, \eqref{det} and \eqref{invmet},
{\tiny\begin{equation} \begin{aligned}
     \partial_i ( \sqrt{\det E} \tilde{C}^{(\partial \phi)^2} G^{ij} \partial_j \Phi) =  \partial_{\rho} ( \sqrt{\det E} \tilde{C}^{(\partial \phi)^2} G^{pp} \partial_{\rho} \Phi) +  \sqrt{\det E_i} \tilde{C}^{(\partial \phi)^2} G_1^{-1}(p) \partial_{x_i} ( \sqrt{\det (i)}  \eta^{x_ix_j} \partial_{x_j} \Phi) \\
     +  \sqrt{\det E_{\mu}} \tilde{C}^{(\partial \phi)^2} \frac{G_2(p)}{G_2(p)^2 - G_1(p)^2} \partial_{\mu} (\sqrt{\det (\mu)} \delta^{\mu \nu} \partial_{\nu} \Phi) + \sqrt{\det E_{\Omega}} \tilde{C}^{(\partial \phi)^2}  G_4(p)^{-1} \partial_{\alpha_i} ( \sqrt{\det(\Omega)} G^{\alpha_i \alpha_j} \partial_{\alpha_j} \Phi),
\end{aligned} \end{equation}}
which can be rewritten as
{\footnotesize\begin{equation} \begin{aligned}
     \partial_i ( \sqrt{\det E} \tilde{C}^{(\partial \phi)^2} G^{ij} \partial_j \Phi) =  \partial_{\rho} ( \sqrt{\det E} \tilde{C}^{(\partial \phi)^2} G^{pp} \partial_{\rho} \Phi) +  \sqrt{\det E} \tilde{C}^{(\partial \phi)^2}  \Delta_{i} \Phi      +  \sqrt{\det E} \tilde{C}^{(\partial \phi)^2}  \Delta_{\mu} \Phi \\
     + \sqrt{\det E} \tilde{C}^{(\partial \phi)^2} \Delta_{\Omega} \Phi).
\end{aligned} \end{equation}}
Substituting \eqref{invmet},
\begin{equation} \begin{aligned}
     \partial_i ( \sqrt{\det E} \tilde{C}^{(\partial \phi)^2} G^{ij} \partial_j \Phi) =  \partial_{\rho} ( \sqrt{\det E} \tilde{C}^{(\partial \phi)^2} G^{pp} \partial_{\rho} \Phi)  \\
     + \sqrt{\det E} \tilde{C}^{(\partial \phi)^2} \left ( G_1^{-1}(p)\Delta_{\eta} \phi   +  \frac{G_2(p)}{G_2(p)^2 - G_1(p)^2} \Delta_{\delta} \phi + G_4(p)^{-1} \Delta_{\Omega} \Phi \right ),
\end{aligned} \end{equation}
and therefore, the equation of motion is
{\small \begin{equation} \begin{aligned}
    \partial_{\rho} ( \sqrt{\det E} \tilde{C}^{(\partial \phi)^2} G_3 \partial_{\rho} \Phi) +      \sqrt{\det E} \tilde{C}^{(\partial \phi)^2} \left ( G_1^{-1}(p)\Delta_{\eta} \phi   +  \frac{G_2(p)}{G_2(p)^2 + B_{23}^2} \Delta_{\delta} \phi + G_4(p)^{-1} \Delta_{\Omega} \Phi \right ) \\
    - \sin 2 \psi \partial_{\rho} (K(p) B_{23}) F_{01} = 0.
\end{aligned} \end{equation}}

\subsubsection*{$\bullet \; A_{5,6,7,8}$}
\label{secaper}

Now, $\delta_{0,s} \delta_{1,s} = 0$, and also $C_{rs}^{\chi F} = C^{\partial \chi F}_{rsi} = 0 $ for $s \neq 2,3$, so
\begin{equation}
    \begin{aligned}
        \frac{1}{4} \partial_r \Bigg[ 
2 \Big (C^{FF}_{rs\,mn}   + \frac{1}{\sqrt{\det E}} ( \rho^2 + l_0^2)^2 \epsilon_{rsmn} \Big) F_{mn}   \Bigg] = 0.
    \end{aligned}
\end{equation}
Substituting $C = \sqrt{\det E} \tilde{C}$, with $\tilde{C}$ calculated in \ref{termsoftheaction}, we obtain the equation of motion,
\begin{equation}
    \begin{aligned}
    \partial_r \Bigg[ 
 \Big (C^{FF}_{rp\,mn}   + \frac{1}{\sqrt{\det E}} ( \rho^2 + l_0^2)^2 \epsilon_{rsmn} \Big) F_{mn} \Bigg] = 0.
    \end{aligned}
\end{equation}
But under these constraints, the second term vanishes (since $mnr$ can only take values on the $s$ above, but $F_{s1 s2} = 0$). So
\begin{equation}\begin{aligned}
    \partial_r ( C^{FF}_{rsmn} F_{mn} ) =0.
\end{aligned}\end{equation}
Finally, $C$ is
\begin{equation} \begin{aligned}
    C_{rsmn} = \sqrt{\det E}\frac{e^{-\phi_0}}{4}  \Big (J_{sr}  \, J_{nm}  - \Big ( J_{nr} J_{sm} - J_{mr} J_{sn} + S_{nr} S_{sm} - S_{mr} S_{sn}  \Big )  \Big ) \\
    =  \sqrt{\det E}\frac{e^{-\phi_0}}{4}  \Big ( S_{mr} S_{sn} -  S_{nr} S_{sm}    \Big )  \Big ),
\end{aligned} \end{equation}
then
\begin{equation} \begin{aligned}
    \partial_r ( \sqrt{\det E} e^{-\phi_0} S_{mr} S_{sn} F_{mn} ) = 0,
\end{aligned} \end{equation}
but $S$ is diagonal
\begin{equation} \begin{aligned}
    \partial_r ( \sqrt{\det E} e^{-\phi_0} S_{rr} S_{ss} F_{rs} ) = 0,
\end{aligned} \end{equation}
and $F_{rs} = \partial_r A_s - \partial_s A_r = - \partial_s A_r$,
\begin{equation} \begin{aligned}
    \partial_r ( \det\sqrt{A} e^{-\phi_0} S_{rr} S_{ss} \partial_s A_r ) = 0.
\end{aligned} \end{equation}
Now, nothing depends on $r$ (it goes from $0$ to $3$), so we can discard terms multiplying the whole sum, and obtain
\begin{equation} \begin{aligned}
    S^{rr} \partial_s \partial_r A_r = 0.
\end{aligned} \end{equation}
So we can still impose the same conditions of eq 52 in \cite{Filev_2007},
\begin{equation} \begin{aligned}
    \partial_1 A_1 - \partial_0 A_0 = \partial_2 A_2 + \partial_3 A_3 = 0,
\end{aligned} \end{equation}
because $S_{00} = -S_{11}, S_{22} = S_{33}$.
\subsubsection*{$\bullet \; A_{2,3}$}
\label{seca23}

\begin{equation}
    \begin{aligned}
    \label{a23}
         \partial_r \Bigg[ 
2 \Big (C^F_{rs\,mn}  \Big) F_{mn}  + C^{\chi F}_{rs} \chi 
\Bigg] = 0
    \end{aligned}
\end{equation}
where, actually, we really have $\partial_r (C^{FF}_{ksmn} G^{sr} F_{mn})$. So we can separate $r$ in three subset
\begin{equation} \begin{aligned}
    r = \rho, \quad    r = \alpha_i \in \Omega_3, \quad
    r = x_0, \dots , x_3 \in D_3,
\end{aligned} \end{equation}
and the first term of \eqref{a23},
\begin{equation} \begin{aligned}
    2 (\partial_{\rho} (G^{pp} C_{psmn} F_{mn}) + \partial_{\alpha_i} (G^{\alpha_i \alpha_j} C_{\alpha_j smn} F_{mn} ) + \partial_{x_i} ( G^{x_i x_j} C_{x_j smn} F_{mn} )).
\end{aligned} \end{equation}
We are going now to obtain for each possible value of $r$, with the same notation of \eqref{E}, \eqref{invmet} and, \eqref{det}, 

$\bullet \; r =p$,

\begin{equation} \begin{aligned}
    C_{psmn} = \sqrt{\det E}\frac{e^{-\phi_0}}{4}  \Big (    S_{mp} S_{sn} -  S_{np} S_{sm}  \Big ) ,
\end{aligned} \end{equation}
\begin{equation} \begin{aligned}
    (\partial_{\rho} (G^{pp} C_{psmn} F_{mn}) = \partial_{\rho} ( G^{pp}  \sqrt{\det E}\frac{e^{-\phi_0}}{2}    S_{pp} S_{ss}F_{ps}).
\end{aligned} \end{equation}

$\bullet \; r = \alpha_i$

\begin{equation} \begin{aligned}
    \partial_{\alpha_i} (G_{\alpha_i \alpha_j} C^{\alpha_j smn} F_{mn} ) = \partial_{\alpha_i} ( G^{\alpha_i \alpha_j} \sqrt{\det E} \frac{e^{-\phi_0}}{2} S_{\alpha_j \alpha_j} S_{ss} F_{\alpha_j s} ),
\end{aligned} \end{equation}
\begin{equation} \begin{aligned}
    \partial_{\alpha_j} (C^{\alpha_j smn} F_{mn} ),
\end{aligned} \end{equation}
\begin{equation} \begin{aligned}
    C_{rsmn} 
    =  \sqrt{\det E}\frac{e^{-\phi_0}}{4}  \Big ( S_{mr} S_{sn} -  S_{nr} S_{sm}    \Big ),
\end{aligned} \end{equation}
so,
{\footnotesize\begin{equation} \begin{aligned}
    \partial_{\alpha_i} (G_{\alpha_i \alpha_j} \sqrt{\det E} \frac{e^{-\phi_0}}{2} S_{\alpha_j \alpha_j} S_{ss} F_{\alpha_j s}) = \partial_{\alpha_i} ( G_{\alpha_i \alpha_j} \sqrt{\det E} \frac{e^{-\phi_0}}{2} \frac{1}{G_{\alpha_j \alpha_j}} \frac{G_{2}}{G_{2}^2+B^2} F_{\alpha_j s}) \\
    =  \partial_{\alpha_i} ( G^{\alpha_i \alpha_j} \sqrt{\det E} \frac{e^{-\phi_0}}{2} \frac{1}{G_{\alpha_j \alpha_j}}   \frac{G_{22}}{G_{2}^2+B^2} \partial_{\alpha_j} A_s) = \partial_{\alpha_j} (G^{\alpha_j \alpha_j} \sqrt{\det(\Omega) \times \kappa_G} \frac{e^{-\phi_0}}{2 G_{\alpha_j \alpha_j}}  \frac{G_{2}}{G_{2}^2+B^2} \partial_{\alpha_j} A_s) \\
    = \sqrt{\kappa_G} \frac{e^{-\phi_0}}{2}  \frac{G_{2}}{G_{2}^2+B^2}  \partial_{\alpha_i} (\sqrt{\det \Omega} \partial_{\alpha_i} A_s) =\sqrt{\kappa_G} \frac{e^{-\phi_0}}{2}  \frac{G_{2}}{G_{2}^2+B^2}  \partial_{\alpha_i} (\sqrt{\det G_{\Omega_3}} G^{ij} \partial_{\alpha_j} A_s) .
\end{aligned} \end{equation}}
Since $A_s$ is a component of a vector, and $G^{ij} = G_4^{-1} G_{\Omega}^{ij}$
\begin{equation} \begin{aligned}
    \sqrt{\det E} \frac{e^{-\phi_0}}{2}  \frac{G_{2} G_4^{-1}}{G_{2}^2+B^2} \Delta_{\Omega_3} A_s .
\end{aligned} \end{equation}
$\bullet \; r  = x_0,x_1$
{\footnotesize\begin{equation} \begin{aligned}
    \partial_{i} (G_{ij} \sqrt{\det E} \frac{e^{-\phi_0}}{2} S_{jj} S_{ss} F_{\alpha_j s}) = \partial_{i} ( G_{ij} \sqrt{\det E} \frac{e^{-\phi_0}}{2} \frac{1}{G_{jj}} \frac{G_{2}}{G_{2}^2+B^2} F_{\alpha_j s}) \\
    =  \partial_{\alpha_i} ( G^{\alpha_i \alpha_j} \sqrt{\det E} \frac{e^{-\phi_0}}{2} \frac{1}{G_{\alpha_j \alpha_j}}   \frac{G_{22}}{G_{2}^2+B^2} \partial_{\alpha_j} A_s) = \partial_{\alpha_j} (G^{\alpha_j \alpha_j} \sqrt{\det(\Omega) \times \kappa_G} \frac{e^{-\phi_0}}{2 G_{\alpha_j \alpha_j}}  \frac{G_{2}}{G_{2}^2+B^2} \partial_{\alpha_j} A_s) \\
    = \sqrt{\kappa_G} \frac{e^{-\phi_0}}{2}  \frac{G_{2}}{G_{2}^2+B^2}  \partial_{\alpha_i} (\sqrt{\det \Omega} \partial_{\alpha_i} A_s) =\sqrt{\kappa_G} \frac{e^{-\phi_0}}{2}  \frac{G_{2}}{G_{2}^2+B^2}  \partial_{\alpha_i} (\sqrt{\det G_{\Omega_3}} G^{ij} \partial_{\alpha_j} A_s) ,
\end{aligned} \end{equation}}
and so,
\begin{equation} \begin{aligned}
     \partial_{x_i} ( G^{x_i x_j} C_{x_j smn} F_{mn} ) = \sqrt{\det E} \frac{e^{-\phi_0}}{2}  \frac{G_{2} G_1^{-1}}{G_{2}^2+B^2} \Delta_{0,1} A_s .
\end{aligned} \end{equation}
But for $x_2, x_3$,
{\small \begin{equation} \begin{aligned}
    \partial_{x_i} (G^{x_j x_j} \sqrt{\det E} \frac{e^{-\phi_0}}{2} S_{x_j x_j} S_{ss} F_{x_j s}) = \partial_{x_i} (G_{x_j x_j} \sqrt{\det E} \frac{e^{-\phi_0}}{2} \frac{G_{x_j x_j}}{G_{x_j x_j}^2 + B^2} \frac{G_{2}}{G_{2}^2+B^2} F_{x_j s}) \\
    = \partial_{x_j} ( \sqrt{\det E} \frac{e^{-\phi_0}}{2} \frac{G_{x_j x_j}^2}{G_{x_j x_j}^2 + B^2} \frac{G_{2}}{G_{2}^2+B^2} \partial_{x_j} A_s) = \sqrt{\det E}  \frac{e^{-\phi_0}}{2} \frac{G_{2}}{G_{2}^2+B^2} \Big (\frac{G_2}{G_2 + B^2}  \Delta_{1,2} A_s \Big).
\end{aligned} \end{equation}}
Now, we add all the terms,
{\small \begin{equation} \begin{aligned}
    \partial_{\rho} ( G^{pp}  \sqrt{\det E}e^{-\phi_0}    S_{pp} S_{ss}\partial_{\rho} A_s) +   \sqrt{\det E} e^{-\phi_0}  \frac{G_{2} G_4^{-1}}{G_{2}^2+B^2} \Delta_{\Omega_3} A_s  +  \sqrt{\det E}e^{-\phi_0}\frac{G_{2} G_1^{-1}}{G_{2}^2+B^2} \Delta_{0,1} A_s  \\
    +  \sqrt{\det E} e^{-\phi_0}\frac{G_{2}}{G_{2}^2+B^2} \Big (\frac{G_2}{G_2 + B^2}  \Delta_{1,2} A_s \Big) + \partial_r ( C^{\chi F}_{rs} \chi) = 0
\end{aligned} \end{equation}}

\subsubsection*{$\bullet \; A_{0,1}$}
\label{sec01}

\begin{equation}
    \begin{aligned}
        \frac{1}{4} \partial_r \Bigg[ 
2 \Big (C^F_{rs\,mn} \Big) F_{mn} +\frac{1}{\sqrt{\det E}} \Bigg (\partial_{\rho} (K(p) B_{23}) (\delta_{0r} \delta_{1s} - \delta_{0s} \delta_{1r}) \sin 2 \psi \Bigg ) \Phi \Bigg] = 0.
    \end{aligned}
\end{equation}
Where, actually, we really have $\partial_r (C^{FF}_{ksmn} G^{sr} F_{mn})$. So we can separate $r$ in three subset again
\begin{equation} \begin{aligned}
    r = \rho, \quad
    r = \alpha_i \in \Omega_3, \quad
    r = x_0, \dots , x_3 \in D_3,
\end{aligned} \end{equation}
obtaining
\begin{equation} \begin{aligned}
    2 (\partial_{\rho} (G^{pp} C_{psmn} F_{mn}) + \partial_{\alpha_i} (G^{\alpha_i \alpha_j} C_{\alpha_j smn} F_{mn} ) + \partial_{x_i} ( G^{x_i x_j} C_{x_j smn} F_{mn} )).
\end{aligned} \end{equation}
So, for

$\bullet \;  r=p$,

\begin{equation} \begin{aligned}
    C_{psmn} = \sqrt{\det E}\frac{e^{-\phi_0}}{4}  \Big (    S_{mp} S_{sn} -  S_{np} S_{sm}  \Big ) ,
\end{aligned} \end{equation}
so
\begin{equation} \begin{aligned}
    (\partial_{\rho} (G^{pp} C_{psmn} F_{mn}) = \partial_{\rho} ( G^{pp}  \sqrt{\det E}\frac{e^{-\phi_0}}{2}    S_{pp} S_{ss}F_{ps}).
\end{aligned} \end{equation}

$\bullet \; r = \alpha_i$

\begin{equation} \begin{aligned}
    \partial_{\alpha_i} (G_{\alpha_i \alpha_j} C^{\alpha_j smn} F_{mn} ) = \partial_{\alpha_i} ( G^{\alpha_i \alpha_j} \sqrt{\det E} \frac{e^{-\phi_0}}{2} S_{\alpha_j \alpha_j} S_{ss} F_{\alpha_j s} ),
\end{aligned} \end{equation}
so
\begin{equation} \begin{aligned}
    \partial_{\alpha_i} (G^{\alpha_j \alpha_j} \sqrt{\det E} \frac{e^{-\phi_0}}{2} S_{\alpha_j \alpha_j} S_{ss} F_{\alpha_j s})   =\sqrt{\kappa_G} \frac{e^{-\phi_0}}{2}  \frac{1}{G_1}  \partial_{\alpha_i} (\sqrt{\det G_{\Omega_3}} G^{ij} \partial_{\alpha_j} A_s) ,
\end{aligned} \end{equation}
but since $A_s$ is a component of a vector,
\begin{equation} \begin{aligned}
    \sqrt{\det E} \frac{e^{-\phi_0}}{2}  \frac{1}{G_1 G_4} \Delta_{\Omega_3} A_s .
\end{aligned} \end{equation}

$\bullet \; r  = x_0,x_1$

\begin{equation} \begin{aligned}
     \partial_{x_i} ( G^{x_i x_j} C_{x_j smn} F_{mn} ) = \sqrt{\det E} \frac{e^{-\phi_0}}{2}  \frac{1}{G_1^2} \Delta_{0,1} A_s ,
\end{aligned} \end{equation}
but for $x_2, x_3$
\begin{equation} \begin{aligned}
    \partial_{x_i} (G^{x_j x_j} \sqrt{\det E} \frac{e^{-\phi_0}}{2} S_{x_j x_j} S_{ss} F_{x_j s})  = \sqrt{\det E}  \frac{e^{-\phi_0}}{2} \frac{1}{G_1} \Big (\frac{G_{2}^2}{G_{2}^2 + B^2}  \Delta_{1,2} A_s \Big).
\end{aligned} \end{equation}
Now, we add to the equation of motion, and use $A_p = 0$,
{\small \begin{equation} \begin{aligned}
    \partial_{\rho} ( G^{pp}  \sqrt{\det E}\frac{e^{-\phi_0}}{2}    S_{pp} S_{ss}\partial_{\rho} A_s) + \sqrt{\det E} e^{-\phi_0}  \frac{1}{G_1 G_4} \Delta_{\Omega_3} A_s  + \sqrt{\det E} e^{-\phi_0}  \frac{1}{G_1^2} \Delta_{0,1} A_s \\
    + \sqrt{\det E}  e^{-\phi_0} \frac{1}{G_1} \Big (\frac{G_{2}^2}{G_{2}^2 + B^2}  \Delta_{3,4} A_s \Big) + \partial_r \Bigg (  \Bigg (\partial_{\rho} (K(p) B_{23}) (\delta_{0r} \delta_{1s} - \delta_{0s} \delta_{1r}) \sin 2 \psi \Bigg ) \Phi \Bigg  ) = 0.
\end{aligned} \end{equation}}

\bibliographystyle{plain} % Escolha o estilo da bibliografia
\bibliography{TsT-penrose} % Nome do seu arquivo .bib (sem a extensão .bib)

\newpage

\end{document}